\documentclass[a4paper,10pt,notitlepage]{report}
\usepackage[usenames,dvipsnames]{color}
\usepackage{geometry}
\usepackage{multirow}
\usepackage{pifont}
\usepackage{graphicx}
\usepackage{threeparttable}
\usepackage{url} 
\usepackage{xcolor,colortbl}
\usepackage[final]{pdfpages}
\usepackage{longtable}
\usepackage[gray]{datapie}
\usepackage{datatool}
\usepackage[gray]{databar}
\usepackage{wrapfig}
\usepackage{scrpage2}
\usepackage[section] {placeins}
\usepackage{ifthen}
\usepackage{wasysym}

\newcommand{\newevenside}{
        \ifthenelse{\isodd{\thepage}}{\newpage}{
        \newpage
        \phantom{placeholder} % doesn't appear on page
        \thispagestyle{empty} % if want no header/footer
        \newpage
        }
}
\newcommand{\op}{\emph{open}EHR}
\newcommand{\opr}{opereffa}

\newcommand{\dm}{dry mouth}
\newcommand{\ar}{archetype}
\newcommand{\arc}{architecture}
\begin{document}
%%%%%%%%%%%%%%%%%%%%%%%%%%%%%%%%%%%%%%%%%%%%%%%%%%%%%%%%%%%% Titelsida
\begin{titlepage}
\begin{centering}
{\sc Thesis for the degree of Licentiate of Engineering}
\vspace{25ex}

{\LARGE\bf Towards Usable \op-aware Clinical Decision Support:}
\vspace{1.5ex}
{\Large \\ A User-centered Design Approach}

\vspace{7ex}

\large Hajar Kashfi
%\vspace{15ex}

\vfill
%\begin{figure}[!h]
%  \centering
%\includegraphics{Pics/eps/CTH-logo.pdf}
%\end{figure}
%\vspace*{-0.7cm}
\emph{Division of Interaction Design}\\
\emph{Department of Applied Information Technology}\\
CHALMERS UNIVERSITY OF TECHNOLOGY\\
Gothenburg, Sweden 2011

\end{centering}
\end{titlepage}

%%%%%%%%%%%%%%%%%%%%%%%%%%%%%%%%%%%%%%%%%%%%%%%%%%%%%%%%%%%% Tryckortssida
{\setlength{\parindent}{0pt}
{\large\bf }
\vspace{1ex}

Towards Usable \op-aware Clinical Decision Support: \\ A User-centered Design Approach
\\
\\
Hajar Kashfi

\copyright Hajar Kashfi, 2011.
\vspace{4ex}

ISSN 1651-4769;7 \\
%Department of Applied Information Technology
%\vspace{4ex}
Department of Applied Information Technology\\
Chalmers University of Technology\\
SE--412~96~~Gothenburg, Sweden\\
Phone: +46 (0)31--772~1000
\vspace{15ex}

{\bf Contact information:}\\
Hajar Kashfi\\
Division of Interaction Design\\
Department of Applied Information Technology\\
Chalmers University of Technology\\
%Rnnvgen 6B\\
SE--412~96~~Gothenburg, Sweden

\begin{tabbing}
Phone: \=+46 (0)31--772~5407\\
Fax:   \>+46 (0)31--772~3663\\
Email: \>{\tt hajar.kashfi@chalmers.se}\\
URL:   \>{\tt http://puix.org}
\end{tabbing}

\vfill
Printed in Sweden\\
Chalmers Reproservice\\
Gothenburg, Sweden 2011
}
\thispagestyle{empty}

\newpage
\thispagestyle{empty}
\vspace*{2cm}

\begin{center}
To my love, \emph{Mohsen} \\
\vspace*{0.2cm}
and my wonderful parents and brothers
\end{center}

\cleardoublepage
\thispagestyle{empty}
\vspace*{3cm}

\newpage
\pagenumbering{gobble}
\vspace*{1.7cm}
%%%%%%%%
{\setlength{\parindent}{0pt}
	{\LARGE\bf Towards Usable \op-aware Clinical Decision Support:  }
	\vspace{1.5ex}
	{\Large \\ A User-centered Design Approach}
	\vspace{3.5ex}
	
	Hajar Kashfi
	
	{\it Department of Applied Information Technology, \\ Chalmers University of Technology}
	
	\vspace{3ex}
	%{Thesis for the degree of Licentiate of Engineering, a Swedish degree\\
	%between M.Sc.\@ and Ph.D.}
	%}
	
	\section*{Abstract}
	%%\chapter*{Abstract}
	%\begin{abstract}
	\vspace*{0.25cm}
	Nowadays, the use of computerized approaches to support health care processes in order to improve quality of health care is widespread in the clinical domain.
	Electronic health records (EHR) and clinical decision support (CDS) are considered to be two complementary approaches to improve quality of health care.
	
	It is shown that EHRs are not able to improve quality of health care without being supported by other features such as CDS.
	On the other hand, one of the success factors of CDS is its integration into EHR, and since there are various international EHR standards (such as \op) being developed, it is crucial to take these standards into consideration while developing CDS. 
	
	Various clinical decision support systems (CDSS) are developed but unfortunately only a few of them are being used routinely. Two of the reasons for unacceptability of CDSSs among their users, i.e.  clinicians, are shown to be their separation from EHRs and poor usability of the user interfaces.
	Besides integration into underlying information framework, i.e.  EHR systems, consideration of human-computer interaction (HCI) in designing and evaluating CDS is one of the success factors that developers of these systems should keep in mind.
	
	This thesis addresses the question of \emph{how  usable \op-aware clinical decision support can be designed and developed in order to improve the quality of health care.} To answer this research question, several sub-questions were identified and investigated. 
	% Applying various theoretical and empirical research strategies in this study
	This included analyzing ``state of the art'' in two different aspects of design and development and evaluation of CDS and also investigating application of a customized user-centered design (UCD) process in developing \op-based clinical applications. 
	
	Analysis of state of the art in interplay between HCI and CDS and also the intersection between CDS and EHR revealed that consideration of both HCI and integration of CDS into EHR is more appreciated in theory than in practice and there is still a long way to go before reaching an acceptable level in these two success factors of CDS.
	
	Moreover, the experience in designing an \op-based clinical application revealed that apart from benefits offered by \op{} approach, such as specifying different roles and involvement of domain experts in defining domain concepts, there are various shortcomings that need to be improved, for instance the limited support for \op{} application developers.
	Additionally, this study revealed that there are characteristics of the domain, tasks and users in the domain that developers should be informed about while applying UCD methods.
	
	%\end{abstract}
	
	\vspace{3ex}
	\noindent {\bf Keywords:} Medical  informatics, Electronic health record, \op, Clinical decision support system, User-centered design, Human-computer interaction,  Interaction design, Usability.

%%%%%%%%
\thispagestyle{empty}

%\newpage
%\vspace*{1cm}
%\thispagestyle{empty}

\cleardoublepage
\thispagestyle{empty}
\vspace*{1.7cm}

\section*{Acknowledgments}
%\chapter*{Acknowledgments}
\vspace*{0.25cm}

I would like to express my gratitude to all of the individuals whose support and help has made this research a reality; especially my supervisor and advisor, Olof Torgersson. My gratitude also extends to G{\"o}ran Falkman, Mats Jontell, Marie Lindgren, Marita Nilsson, and the members of the Clinic of Oral Medicine at Sahlgrenska University Hospital who were involved in this project.
% as interviewees or test users.

I also want to thank Sus Lundgren, Martin Hjulstr{\"o}m, Erik Fagerholt and Soren Lauesen who provided constructive feedback on the design work presented in this thesis. I am also indebted to Marie Gustafsson Friberger who provided valuable comments on the thesis. Moreover, I appreciate the opportunities for helpful discussions provided by the openEHR community members, especially Ian McNicoll, Koray Atalag, Pablo Pazo, Rong Chen, Seref Arikan, and others whose names may have been omitted here.

I would also like to convey my gratitude to my colleagues in the Interaction Design Division, in particular Staffan Bj{\"o}rk and Fang Chen, and members of the Human-Technology Design research school of which I too am a member. Of course, I would be remiss not to mention my kind fellow graduate student, Anna Gryszkiewicz, with whom I share a pleasant and peaceful office.
% and who is always willing to engage in conversations that often lead to moments of inspiration and clarity.

My friends know that they all hold a very special place in my heart and I am grateful to them for all of the joy they have brought into my life over the past few years. But my wonderful family--my parents and brothers--know that they are and always will be a part of my heart. I hope that the realization that their love, patience, and willingness to bear the burden of our separation, has empowered me to make my dreams come true will allow them to look upon my achievements here with pride.

And finally, thanks to you, \emph{Mohsen}, my love, and the meaning of my life. You are the best and happiest thing that has ever happened to me. You are not only a marvelous life partner, but also a tremendous friend and a remarkable fellow graduate student. I appreciate the fruitful discussions we have had together and the feedback you provided on this thesis. I am grateful to you for being such a cheerful and supportive soul mate through both sunshine and rain, as well as all of the other moments that will forever remain between you and I. Thank you for being my everything and everyone--your love is the axis of my entire being.

\vspace*{0.8cm}

\begin{footnotesize}
\begin{flushright}
Hajar Kashfi\\
Gothenburg, May 2011
\end{flushright}
\end{footnotesize}

\cleardoublepage
\vspace*{3cm}
\thispagestyle{empty}

\newpage
\renewcommand{\thepage}{\roman{page}}
\setcounter{page}{1}
\vspace*{1.7cm}
\section*{List of Appended Papers}
\vspace*{0.25cm}
%\chapter*{List of Appended Papers}

\noindent This thesis is a summary of the following four
papers. References to the papers will be made using the Roman numbers
associated with the papers.

\newcounter{Lcount}
\begin{list}{\Roman{Lcount}}{\usecounter{Lcount}}
 
  \item{\textbf{Hajar Kashfi},
      ``\emph{Towards Interaction Design in Clinical Decision Support Development: A Literature Review},''
       submitted to {\it International Journal of Medical Informatics}, Elsevier}
   
  \item{\textbf{Hajar Kashfi},
      ``\emph{The Intersection of Clinical Decision Support and Electronic Health Record: A Literature Review},''
       submitted to {\it 1$^\mathrm{st}$ International Workshop on Interoperable Healthcare Systems (IHS’2011) - Challenges, Technologies, and Trends}, Szczecin, Poland, September 19-21, 2011.}
      
  \item{\textbf{Hajar Kashfi},
      ``\emph{Applying a User Centered Design Methodology in a Clinical Context},''
      in {\it  Proceedings of The 13$^\mathrm{th}$ International Congress on Medical Informatics (MedInfo2010), Studies in health technology and informatics}, 2010 Jan ;160(Pt 2):927-31.}
  
  \item{\textbf{Hajar Kashfi},
      ``\emph{Applying a User-centered Approach in Designing a Clinical Decision Support System},''
       submitted to {\it Computer Methods and Programs in Biomedicine}, Elsevier.}
      
    \item{\textbf{Hajar Kashfi}, Olof Torgersson,
      ``\emph{Supporting \op{} Java Desktop Application Developers},''
      to appear in {\it The XXIII  International Conference of the European Federation for Medical Informatics, Proceedings of Medical Informatics in a United and Healthy Europe (MIE2011)},Oslo, Norway, 28-31 August, 2011.} 
     
    \item{\textbf{Hajar Kashfi},  Jairo Robledo Jr.,
      ``\emph{Towards a Case-Based Reasoning Method for \op-Based Clinical Decision Support},''
       to appear in {\it Proceedings of The 3$^\mathrm{rd}$ International Workshop on Knowledge Representation for Health Care (KR4HC'11)}, Bled, Slovenia, 6  July, 2011.}  
\end{list}

%%%%%%%%%%%%%%%%%%%%%%%%%%%%%%%%%%%%%%%%%%%%%%%%%%%%%%%%%%%% Other papers
\newpage
\vspace*{1.7cm}
\section*{List of Other Papers}
\vspace*{0.25cm}
%\chapter*{List of Other Papers}
%
%\noindent This thesis is a summary of the following four
%papers. References to the papers will be made using the Roman numbers
%associated with the papers.

\newcounter{LLcount}
\begin{list}{\Roman{LLcount}}{\usecounter{LLcount}}
 
  \item{\textbf{Hajar Kashfi},
      ``\emph{An openEHR-Based Clinical Decision Support System: A Case Study},'' 
      in {\it The XXII International Conference Of The European Federation For Medical Informatics, Proceedings of Medical Informatics in a United and Healthy Europe (MIE2009), Studies in health technology and informatics}. 2009. p. 348.}
     
  \item{\textbf{Hajar Kashfi} and Olof Torgersson,
      ``\emph{A Migration to an openEHR-Based Clinical Application},''
      in {\it The XXII International Conference Of The European Federation For Medical Informatics, Proceedings of Medical Informatics in a United and Healthy Europe (MIE 2009), Studies in health technology and informatics}. 2009. p. 152-6.}
\end{list}

\newpage
\tableofcontents
%\listoffigures
%\listoftables
%------------------------------------------------------------------------- 
%------------------------------------------------------------------------- 
%\chapter[Usable \op-aware CDS]{Towards Usable \op-aware Clinical Decision Support} 
%-------------------------------------------------------------------------
\cleardoublepage
\pagenumbering{gobble}
\part{Introduction}
\pagenumbering{arabic}

\renewcommand{\thepage}{\arabic{page}}
\setcounter{page}{2}

\cleardoublepage
\vspace*{3cm}
\thispagestyle{empty}

%\aliaspagestyle{title}{empty}
%\renewcommand*{\titlepagestyle}{empty}

\chapter{Introduction}
\label{ch:intro}
%ERRORS % 17- Errors in clinical care

Errors that occur in a clinical process are mostly due to cognitive limitations of humans, the potential to forget knowledge in the health care flow, and difficulties in clinical workflows \cite{Bates2003a, Kushniruk2004}.
Clinicians are prone to making errors, especially because of the limitations of the working memory \cite{Walker2006}.

Various avoidable errors or adverse events in health care are documented in the literature.
These errors and events have even lead to patients' deaths in some cases.
About 12 preventable deaths per million inhabitants in Sweden are reported \cite{Reizenstein1987}.
According to the Swedish Parliament, around 100,000 patients suffer from preventable adverse clinical events each year and 3000 of these adverse events lead to patient deaths \cite{Parlement2010}.
Preventable medical errors resulted in deaths of up to 98,000 people in the United States in 1999 \cite{Kohn1999}. Around 11\% of the patients admitted to two hospitals in London, experienced adverse events; 48\% of which  were preventable and 8\% of which resulted in patient deaths \cite{Vincent2001}.

There have been investigations about the quality of care in various countries. 
%Serious and widespread problems related to the quality of health care reported in the United States as well as other countries such as Sweden. 
%In 1991, it was assumed that around 5\% of the national Swedish doctors order unnecessary medication, or cause some complications \cite{Reizenstein1991}. 
%The American adults receive around 50\% of the expected health care as reported in \cite{McGlynn2003}.
The existing quality problems belong to three different categories, namely ``underuse'' (failure to provide the best expected health care service), ``overuse'' (providing a health care service that is more harmful than being beneficial for the patient), or ``misuse'' (unsuccessful delivery of the best expected health care service because of some preventive complications), which occur in both small and large health organizations \cite{Chassin1998}. 

Obviously, there are many debates surrounding the quality of health care, but one should start with defining the meaning of ``quality'' in this domain.
A widely accepted and robust definition of quality is the definition developed in 1990 by the Institute of Medicine (IOM) as
``the degree to which health services for individuals and populations increase the likelihood of desired health
outcomes and are consistent with current professional knowledge'' \cite{IOM2010}. 
According to Graham \cite{Graham1995} ``quality must be judged as good if care, at the time it was given, conformed to the practice that could have been expected to achieve the best results.'' 

It has been demonstrated that information systems have the ability to decrease avoidable clinical errors by supporting clinicians in health care process  or in other words to improve the quality of care \cite{Graham2008}. \emph{Medical informatics} is the research field that is dealing with this matter.
Electronic health records (EHR) have been the leading research focus in this field so far \cite{Hasman1997, Greenes2007, Osheroff2005}. The EHR research field deals with issues such as capturing, storing, retrieving and sharing patient data.

For EHRs to be able to improve quality of care, they should be supported by clinical decision support (CDS) services \cite{Sittig2008, Osheroff2005, Greenes2004}, those services that aid clinicians in the process of decision making.
Nonetheless, not all CDS developments have led to improving the clinical practice \cite{Kawamoto2005}.
Hunt et al. \cite{Hunt1998} indicate that 66\% of the CDS implementations have led to significant improvement in health care while the remaining 34\% did not.
Various efforts have been made in order to identify barriers to low adoption, acceptability and ineffectiveness of CDS . Efforts have also been made to identify success factors in developing them \cite{Greenes2007, Kawamoto2005, Bennett2003, Trivedi2002, Osheroff2007}. 
Some of those success factors are the level of integration with clinicians' workflow \cite{Greenes2007, Trivedi2002, Kawamoto2005, Anderson1999, Wetter2002, Berner2007}, the degree of patient-specificity \cite{Greenes2007,Kawamoto2005,Osheroff2007}, availability at the point of care or timely access \cite{Greenes2007, Kawamoto2005}.
Accordingly, two of the main factors that have a direct relation to success of the CDS are integration of the CDS to EHR systems and proper design of CDS by taking \emph{human-computer interaction} (HCI) into consideration.
%It should be considered that although CDSS:s are meant to reduce clinical errors, because of improper design of those systems, other kinds of errors may occur by using them \cite{Graham2008, Saleem2005, Kushniruk2004, Kushniruk2005,Ornstein2003, Koppel2005, Yy2005}. 

While there have been various recommendations regarding consideration of these two success factors in development of CDSSs, related literature suggests that these factors are being partially or totally ignored in many of the projects.
%-------------------------------------------------------------------------
%-------------------------------------------------------------------------
\section{Overview}

This thesis investigates the question of \emph{how usable \op-aware clinical decision support can be designed and developed in order to improve the quality of health care.} In order to answer this question, several sub-problems were identified to be investigated. The study has been carried out in oral medicine. However, the outcome of the research should be applicable to medical informatics in general.
The structure of the thesis is as follows. 
The frame of reference of this research is presented in Chapter~\ref{ch:frame}. This chapter includes the definition of different concepts basic to this research namely medical informatics, EHR, \op, CDS, HCI, usability and user-centered design (UCD).
%The background and related works are discussed in Section~\ref{ch:back}. This includes the efforts made in order to improve quality of health care, and how this relates to the concept of EHR and CDS. The success factors and difficulties in developing CDS and also recommendations and guidelines to develop highly adaptable CDS is also demonstrated in this section.

The research process and the conceptual framework are discussed in Chapter~\ref{ch:process}. 
% ( the research or the answer to the question of \emph{where do we stand in the research field of clinical decision support?} or from a broader perspective \emph{where do we stand in the road to improve the quality of health care as the ultimate goal of medical informatics?}) 
%
The research questions and  objectives of this study are presented in this chapter as well.
Methods and tools are introduced in Chapter~\ref{ch:methods}.
A summary of the included papers is given in Chapter~\ref{ch:sum}, along with how they can be put in relation to each other and the research questions.
%Thesis research contribution is presented in Chapter~\ref{ch:cont} by answering some of the research questions.
Some directions for future work are discussed in Chapter~\ref{ch:fut}.
Finally, a conclusion is provided in Chapter~\ref{ch:conc}.
%------------------------------------------------------------------------- 
%-------------------------------------------------------------------
\chapter{Frame of Reference}
\label{ch:frame}
%-------------------------------------------------------------------
\section{Medical Informatics}
%-------------------------------------------------------------------------
Medical informatics is defined as a scientific discipline ``concerned with the systematic processing of data, information and knowledge in medicine and health care'' \cite{Hasman1996, Haux1997}.  This domain covers both ``computational'' and ``informational'' aspects of the processes in the clinical domain \cite{Haux1997}. 
Medical informatics deals with providing solutions for problems of data, information and knowledge processing in medicine and health care \cite{Hasman1996}.
As a discipline, medical informatics has been around for more than 50 years \cite{Hasman1997} but still is called young especially compared to other medical disciplines \cite{Haux2010}.
Nowadays, new names are suggested for this discipline such as \emph{health informatics} and \emph{clinical informatics} since the word ``medical'' does not cover nursing informatics, dental informatics and so on \cite{Hasman1997}.

%Tasks and aims of the research field of medical informatics are identified as: (1) diagnostic (2) therapy (3) therapy simulation (4) early recognition and prevention (5) compensating physical handicaps (6) health consulting (7) health reporting (8) health care information systems (9) medical documentation (10) comprehensive documentation of medical knowledge and knowledge-based decision support \cite{Haux1997}.
There are various research areas in the field of medical informatics namely electronic health record (EHR) systems, information systems, decision support systems, and image and signal processing \cite{Hasman1997}. EHRs have been the leading research focus in this field so far \cite{Hasman1997, Greenes2007, Osheroff2005}. The EHR research field deals with issues such as capturing, storing, retrieving and sharing patient data. This has led to a number of benefits such as reduced number of transportation errors, higher legibility of reports, and avoiding redundancy \cite{Greenes2007}.
These benefits indirectly affect patient safety, health care quality and efficiency \cite{Greenes2007}.
Recently, there has been more and more interest in adoption of EHRs and developing clinical applications based on EHRs \cite{Greenes2007, Sittig2008, Osheroff2005}.

%-------------------------------------------------------------------------
The idea of benefiting from computers and information technology (IT) in the clinical domain has been around since the 1950s \cite{Collen1986} (or the 1960s as observed by \cite{Greenes2007}) when there were initiatives to automate health care and to simulate the clinical decision making by computers \cite{Greenes2007, Berner2007}.
One of the turning points of medical informatics is considered to be around 50 years ago when in 1959, Ledley and Lusted reported on reasoning foundations of medical diagnosis \cite{Ledley1959}.

Even though the concept of atomization of health care and application of computers and IT in this domain is an old trade,  it has a slow adoption pace and low impact level compared to other domains such as engineering, marketing, etc. In other words, health care has fallen behind other disciplines in applying information technology to improve the processes and outcomes \cite{Greenes2007, Osheroff2005}.

%%-------------------------------------------------------------------------
%There have been several activities in local, national and international level that encourage implementing cutting edge clinical applications \cite{Sittig2008}.
As mentioned above, since the introduction of computers in the clinical domain in the past decades, the main progress in this area has been in coping with  information management, i.e. adopting EHRs \cite{Greenes2007, Osheroff2005, Hasman1997} rather than adopting CDS in order to improve the decision making process.
One of the aims of the efforts in the area of EHR has been to improve quality of health care but it is doubted whether EHRs have the ability to fulfill this goal \cite{Sittig2008}.  More information about EHRs is presented in the following section.

%-------------------------------------------------------------------------
\section{Electronic Health Record}

The idea of computerized medical records has been around as one of the key research areas in medical informatics for more than 20 years.
Iakovidis \cite{Iakovidis1998} defines an EHR as ``digitally stored health care information about an individual's lifetime with the purpose of supporting continuity of care, education and research, and ensuring confidentiality at all times''.
EHRs include the whole range of patient-related data such as demographic information, medical history, medications, and allergies \cite{Blobel2006}. 

The main aim of EHRs is to make distributed and cooperating health information systems and health networks come true \cite{Blobel2006}. 
The first benefit of deploying EHRs is that patients' information is no longer on a piece of paper and clinicians have access to all patients' information when required \cite{Greenes2007}.

Since the introduction of EHRs, various projects were initiated that led to development of various commercial EHR products. Nowadays, there are more and more EHR systems being developed. The interest is also increasing at the governmental level in different countries such as the UK and Sweden. 
However, the EHRs adoption rate is still low in community hospitals and office practices, while higher in academic medical centers \cite{Greenes2007}.
The maximum adoption of EHRs in The United States is demonstrated to be only 40\%  \cite{Ash2005}. In those countries in which there exists a national health care plan, this rate is considered to be higher \cite{Greenes2007}.
Several reasons have been identified for the low adoption rate of EHRs in small hospitals and office practices, viz. high implementation and maintenance costs, additional time and effort and finally the difficulty in choosing among available systems in the market due to a lack of standardization \cite{Greenes2007}.

\subsection{The need for Interoperability}

To to be able to fully benefit from EHRs, timely and secure access to all of the EHR systems should be ensured, EHRs should be up-to-date and accurate in terms of information they contain, and they should be correctly understood when being communicated \cite{Stroetmann2009}. This means that EHR systems should be interoperable.
EHRs are stored in various formats in different products which yields interoperability problems in the domain. Therefore, developing national and international EHR standards is important to support interoperability \cite{Schloeffel2006}.
%\begin{itemize}
%\item ``share patient health information between health professional in a multi-disciplinary shared-care environment''
%\item ``support interoperability between organization within an enterprise, a regional or national health system or in future, across national borders''
%\item ``support interoperability between software from different vendors''
%\end{itemize}

Before going into details of the approaches suggested to enhance interoperability of EHRs, it is proper to present a definition of interoperability. 
Interoperability of health systems is defined as ``the ability, facilitated by ICT applications and systems, to exchange, understand and act on citizens/patients and other health-related information and knowledge among linguistically and culturally disparate health professionals, patients and other actors and organizations within and across health system jurisdictions in a collaborative manner'' \cite{Stroetmann2009}.
Four levels of interoperability are defined by  Stroetmann \cite{Stroetmann2009}: 
\begin{enumerate}
	\item having no interoperability
	\item technical and syntactical interoperability
	\item partial semantic interoperability
	\item full semantic interoperability
\end{enumerate}

A challenging aim regarding EHRs has been to reach semantic interoperability in EHR systems. Interest in this issue in particular is increasing in the EU recently with the aim of reaching semantic interoperability at regional, national and even the EUR level \cite{Stroetmann2009}.

So far, several efforts have been made to develop EHR standards in order to structure and exchange patient information and to enable semantic interoperability among medical information systems. The main approaches are as follows: 
\begin{itemize}
	\item The European Committee for Standardization (CEN) Electronic Health care Record communication standard (CEN/ISO EN13606)
	\item The Governmental Computerized Patient Record project
	\item The Health Level 7 (HL7)  Reference Information Model and its clinical document architecture
	\item The GEHR approach
	\item The \op{} approach which is a continuation to GEHR
\end{itemize}
All the above approaches focus on the technical issues related to standardized and interoperable EHRs.
More information about these approaches is provided in the following sections.
The EHR interoperability standard that is applied in this study is \op{}. 
According to \op{} website \footnote{\url{http://www.openehr.org}} ``the Swedish government has decided on the use of ISO 13606 as a base standard for national health data communication. \op{} will be used to define clinical models, terminology integration, and to implement 13606 in some contexts.'' ISO/CEN 13606 resembles the \op{} reference model (See Section~\ref{sec:rm}) in a simplified manner \cite{oph2009} (ISO/CEN 13606 is explained in Section~\ref{sec:cen}).
This has been a huge motivation for us to consider \op{} as our EHR approach to carry out this study.

%%-----------------------------
%-------------------------------------------------------------------
\section{\op}
\label{sec:oph}

\op{} has its origins in 1992 in an EU research project named Good European Health Record. This project was later continued under the name of Good Electronic Health Record (GEHR) \cite{Bird2003}. Currently the maintenance of \op{} is done by a non-profit organization named the \op{} Foundation \cite{oph2009}. 

In the \op{} approach, clinicians are in charge of defining the specifications of clinical knowledge to be used in information modeling. The main emphasis of \op{} is on semantic interoperability of medical records.
This approach suggests a two-level architecture for clinical applications to separate knowledge and information levels in order to overcome the problems caused by the ever-changing nature of clinical knowledge. Patient data is stored in a generic form which is retrievable in any system using constraints named \emph{\ar s}. An \ar, which is designed by domain experts, defines some constraints on data in terms of types, values, relation of different items and so on. Archetypes are used for data validation and sharing \cite{Beale2009}.

The \op{} framework consists of the reference information model (RM), the implementation technology specification, the archetype definition language (ADL), the open-source implementations, and an archetype repository (the repository is explained more in Section~\ref{sec:arch-dev}) \cite{Beale2009}. A review of the \op{} architecture is presented by Beale \cite{Beale2009}. The key concepts of the \op{} architecture are explained in the following sections.
%-------------------------------------------------------------------
\subsection{Two-level Modeling}

\op{} suggests a \emph{two-level} \arc{} for EHR systems, and accordingly a two-level software engineering approach for developing such systems.
The key idea in the two-level \arc{} is the separation of the domain knowledge level and the  information level. 

The first level or the lower level is a stable reference information model. Software and data are built from this stable object model named the \op{} Reference Model (RM). The second or upper level provides formal definitions of the clinical domain concepts. This  reduces the dependency of the system and underlying data to the ever-changing clinical concepts.
%-------------------------------------------------------------------
\subsection{Two-level Software Engineering}
\label{sec:rm}
%As opposed to the traditional software engineering approach i.e. analysis, design, implementation, etc. which is based on requirement engineering via discussing with users, 
\op{} suggests a two-level software engineering approach. 
In this approach, there exists different view points and a separation of responsibilities in software development. 
The main roles involved in the \op{} process are domain experts, users and IT developers. Different view points introduced by \op{} are the domain knowledge environment, the runtime system and the technical development environment.
The \op{} approach consists of the following steps:
\begin{itemize}
	\item Domain specialists build reusable archetypes, templates (collections of \ar s, see Section~\ref{sec:arc}) for local use and terminologies for general use.
	\item IT developers focus on generic aspects of the system such as data management.
	\item The user works with a GUI which is derived from the templates. Data is generated by users via the EHR system and is validated by \ar s at runtime.
\end{itemize}
%-----------------------Two level SWE------------------------------------------------- 
\begin{figure}
	\centering
	\includegraphics[width=\textwidth]{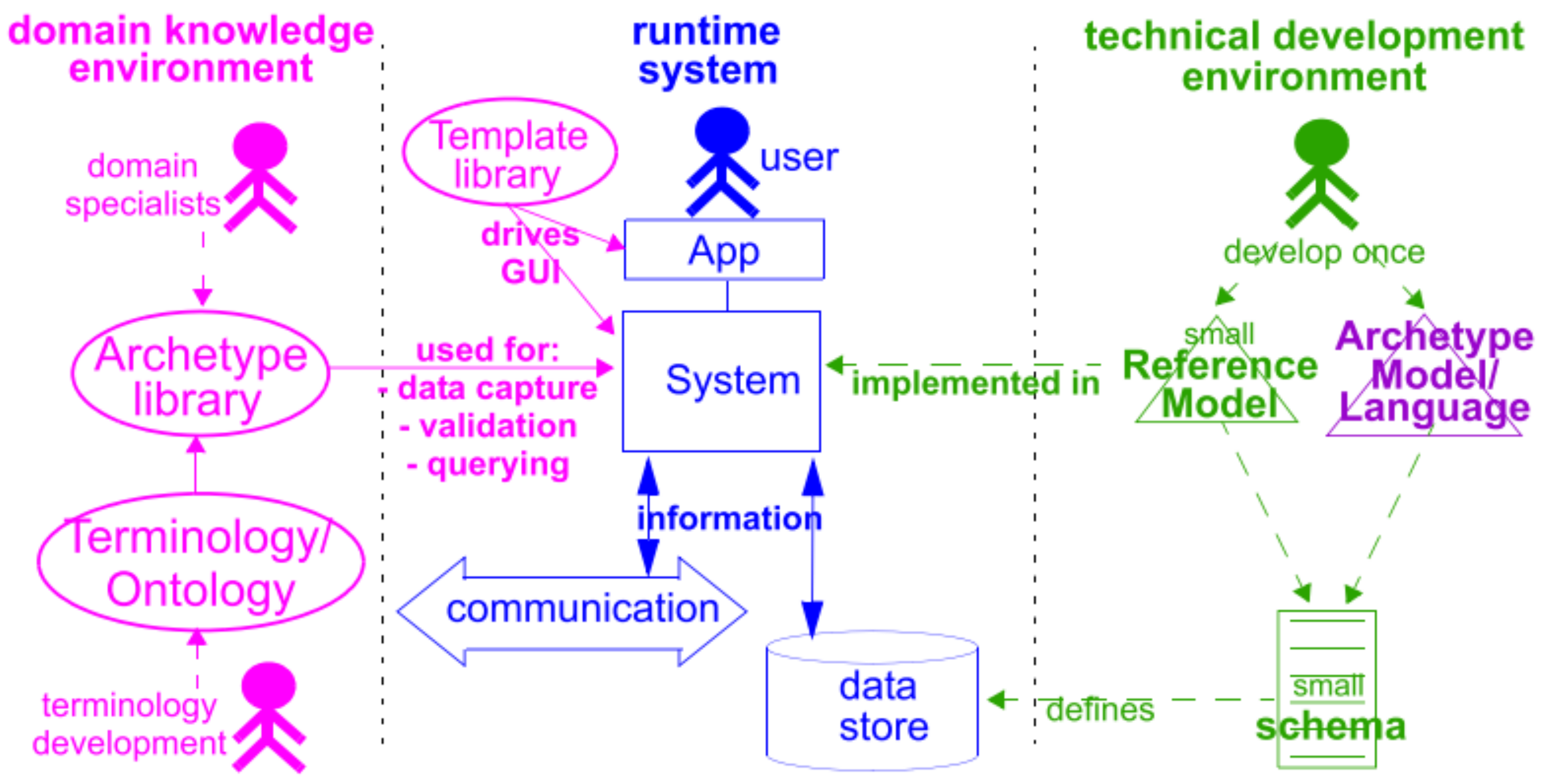}
	\caption{The \op{} two level software engineering taken from \cite{Beale2009}}
	\label{fig:swe}
\end{figure}
%------------------------------------------------------------------------- 
%-------------------------------------------------------------------
\subsection{The Reference Model}

The \op{} RM is specific to the clinical domain but still includes very generic clinical concepts such as composition, observation, evaluation, instruction and action. Moreover, in this RM, different data types are defined such as coded text, quantity and multimedia.
%-------------------------------------------------------------------
\subsection{Archetype}
\label{sec:arc}
In the upper level of the \op{} two-level modeling approach, domain level definitions are defined in form of \ar s and templates. These definitions are used in the EHR system at runtime.  
Archetypes are used to define constraints on the generic RM.
For instance, a blood pressure measurement can be defined in form of an \ar{} in contrast to a more general clinical concept such as an observation which is the focus of the RM.

All the information that is based on the RM is \ar able, or in other words, can be controlled by \ar s in terms of its creation, modification and so on.
Each \ar{} is an instance of the archetype model (AM) and is stored separated from data in an \ar{} repository. The archetype definition language (ADL) is the language that is used to define \ar s based on the AM.
It is recommended that when possible, \ar s should be reused and/or customized instead of being created from scratch. 
The relation of \ar s to the RM is depicted in Figure~\ref{fig:meta}.
%-------------------------------------------------------------------
\subsection{Template}

Templates encapsulate a group of \ar s to be applied for a local use. Templates are trees of one or more \ar s and correspondent to user interface forms, printed reports or other realizations of clinical data \cite{Beale2009}.
For instance, using a template, one can put different clinical concepts like ``blood pressure measurement'' and ``mouth examination´'' (both defined as  \ar s) together to  create an output report for EHRs \cite{Beale2009}.
%----------------------------------------------------------------------------------------------------
%-----------------------Archetype meta-architecture------------------------------------------------ 
\begin{figure}
	\centering
	\includegraphics[width=\textwidth]{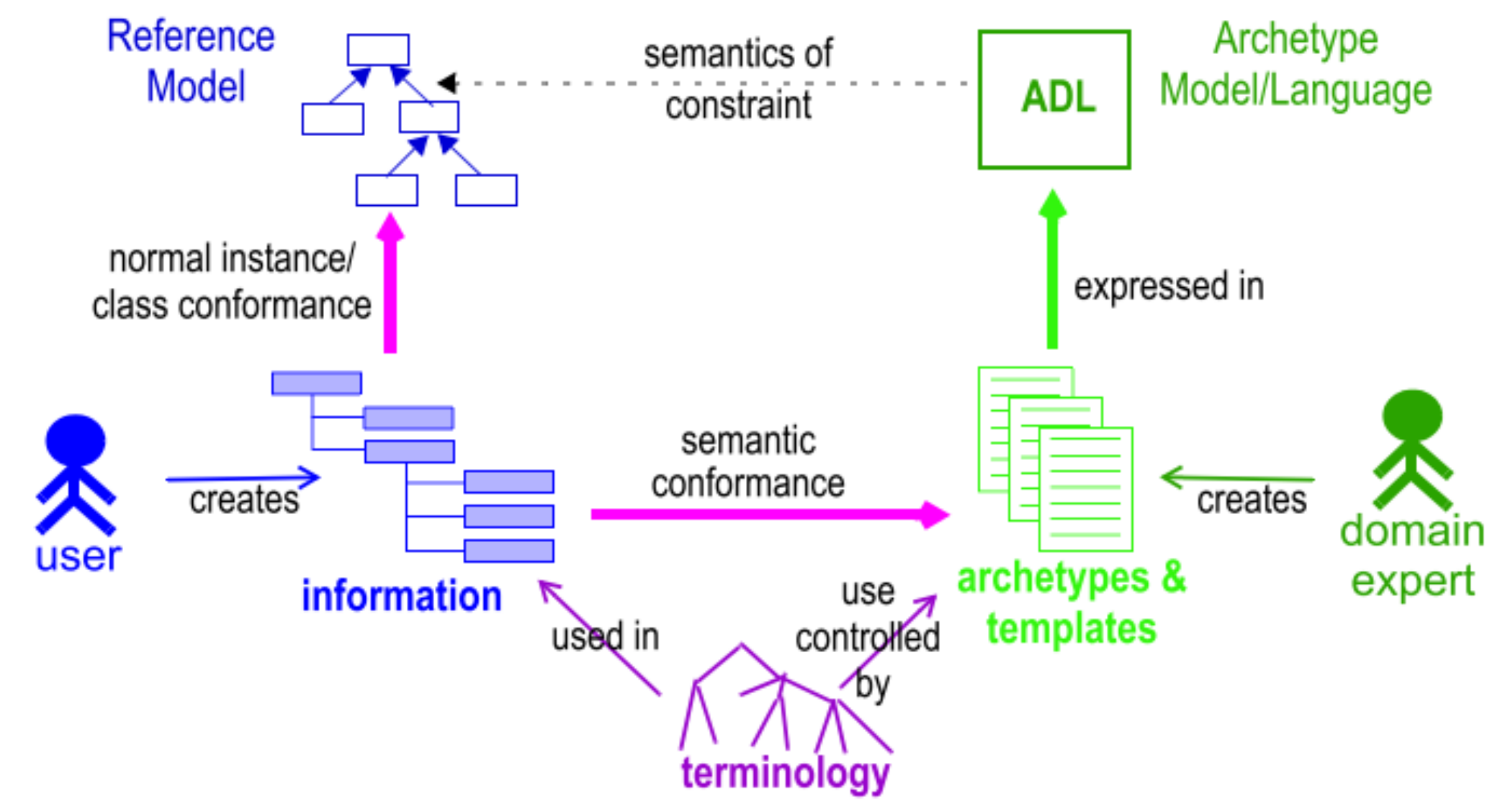}
	\caption{The \op{} \ar{} meta-architecture taken from \cite{Beale2009}}
	\label{fig:meta}
\end{figure}
%------------------------------------------------------------------------- 
%----------------------------------------------------------------------------------------------------
\section{Other EHR Standardization Approaches}
\label{sec:standards}

Several studies have compared the main interoperability standards and specifications in the  clinical domain \cite{Blobel2006,Schloeffel2006}.
Based on a survey, Eichelberg provides information on the most relevant EHR standards \cite{Eichelberg2005}. This includes the level of interoperability provided by each standard, as well as content structure, access services, multimedia support and security.
In the following sections, a brief introduction to the most relevant EHR standardization approaches is presented.
%-------------------------------------------------------------------
\subsection{The CEN/ISO EN13606 standard}
\label{sec:cen}

The CEN/ISO EN13606 is a European norm from CEN which is also approved as an international ISO standard \cite{CEN-web}.
The aim of this standard is to enable semantic interoperability in the electronic health record communication.
The CEN standard is actually a subset of the \op{} specification \cite{Schloeffel2006}.
The same as \op, this standard is based on the idea of a two-level architecture (i.e. a dual model architecture \cite{CEN-web}) which consists of a reference model and \ar s.
%This standard represents a two-level modeling approach the same as the one suggested by the \op{} standard. The CEN standards consists of four parts: ``extended architecture, domain term list, distribution rules, and messages for the exchange of information'' \cite{Blobel2006} 
%-------------------------------------------------------------------
\subsection[The Governmental Computerized Patient Record Project (G-CPR)]{The Governmental Computerized Patient Record \\Project (G-CPR)}

G-CPR is a joint project between the US Department of Defense (DoD), the US Department of Veterans Affairs (DVA) and the Indian Health Services (IHS) \cite{Blobel2006}. This solution uses object-oriented specification to enable interoperability and is rather a service-oriented solution than an architecture-based solution \cite{Blobel2006}. 
%-------------------------------------------------------------------
\subsection{Health Level 7 (HL7)}

HL7 is a well-known EHR communication standard in the clinical domain \cite{Blobel2006}. 
According to the HL7 website\footnote{\url{http://www.hl7.org/}}: ``Health Level Seven International (HL7) is a not-for-profit, ANSI-accredited standards developing organization dedicated to providing a comprehensive framework and related standards for the exchange, integration, sharing, and retrieval of electronic health information that supports clinical practice and the management, delivery and evaluation of health services''.
HL7 in HL7 version 3 presents a comprehensive Reference Information Model (RIM) \cite{Blobel2006}. The HL7 clinical document architecture (CDA) templates are similar to \op{} \ar s \cite{Eichelberg2005}. This standard, provides data-level interoperability but functional level interoperability is not provided \cite{Blobel2006}. 
%-------------------------------------------------------------------------------------------------
%----------------------------------------------------------------------------------------------------
%----------------------------------------------------------------------------------------------------
%----------------------------------------------------------------------------------------------------
%----------------------------------------------------------------------------------------------------
%----------------------------------------------------------------------------------------------------
\section[Decision Support]{Decision Support in the Clinical Domain}

Clinical decision support is a sub-domain of a more general research area named decision-making support. 
According to Gupta \cite{Gupta2006} ``decision-making support systems (DMSS) are \emph{Information Systems} designed to interactively support all phases of a user's decision-making process.''
There are various definitions for  CDSS and CDS in the literature three of which are presented here:

\begin{itemize}
	\item ``computer-based clinical decision support (CDS) can be defined as the use of the computer to bring relevant knowledge to bear on the health care and well being of a patient'' \cite{Greenes2007}.
	\item ``clinical decision support refers broadly to providing clinicians or patients with clinical knowledge and patient-related information, intelligently filtered, or presented at appropriate times, to enhance patient care'' \cite{Osheroff2005}.
	\item ``clincial decision support is any EHR-related process that gives a clinician patient-related healthcare info with the intent of making the clinician's decision making more efficient and better informed'' \cite{Walker2006}.
\end{itemize}
%
%Many references use these terms to refer to a computer program that is supposed to facilitate the process of decision making for decision makers. This support is done by mapping or compiling existing data to useful information that can be used as a clue for making the best decision \cite{Vikram2009}.
CDS impacts the process of decision making about individual patients. This support should be provided at the point of care and while the decisions are made \cite{Berner2007}.
These systems provide support for diagnosis of diseases, prevention of errors in the clinical process, treatment, and future evaluation of the patient.
Services supported by CDS include diagnosis, alerting, reminding, treatment suggestions, and patient education.
CDS interventions are the CDS content and the method for delivering the content.
%
%Berner et al in \cite{Berner2007} specify different types of CDSSs based on some features; such as timing in providing the support (before, during or after the clinical decision is made), and activeness or passiveness of the system to respond to the input provided by the clinician (do the system needs invocation by user or does it provide the support automatically at the point of care based on the input?)

In providing CDS, three modes of interaction between human and computer can be defined \cite{Greenes2007}:
\begin{itemize}
	\item User in charge (users can override computer's suggestions)
	\item Computer in charge (any decision made by computer is expected to be carried out by  users)
	\item Collaborative decision making (computer controls the input, based on users' entries options are provided, users makes the desired choice)
\end{itemize} 

The idea of having both computers and humans in charge of the health care process is more practical in the clinical domain than building intelligent autonomous systems that are in charge of the decision making process \cite{Greenes2007, Berner2007}. The latter may work in other disciplines but is less applicable in the clinical domain.
Berner \cite{Berner2007} discusses that CDS is not meant to come up with ``the answer'' but should provide information for the user and aid him/her in making decisions.

Not all of the information in a clinician's mind can be transfered to the computer (the CDSS) so a clinician usually knows more about the patient. Therefore, having a collaborative pattern in which a clinician can eliminate some of the choices made by the computer is better \cite{Berner2007}.

In this thesis, CDSS and CDS are used interchangeably  to refer to a computer program that aids clinicians in the process of decision making, at the point of care, and based on health information of an individual patient, by presenting that information coupled with external knowledge in a way that is more suitable for making decisions regarding the care process of that specific patient. The system is not meant to make the decisions, rather it is the clinician who makes the final decision.

%-----------------------------------------------------------------------------
\section{Human-Computer Interaction}
\label{sec:hci}

Human-computer interaction (HCI) is defined as ``a discipline concerned with the design, evaluation and implementation of interactive computing systems for human use and with the study of major phenomena surrounding them'' \cite{Hewett1996}.
The concept of usability is considered to be the heart of HCI \cite{Helander1998}. Usability is defined as ``the extent to which a product can be used by specified users to achieve specified goals with effectiveness, efficiency and satisfaction in a specified context of use'' \cite{ISO9241}.
%---------Why Usability--------------------------------------------------------------- 
A great deal of effort in the field of HCI is aimed at designing and developing more usable computer systems \cite{Helander1998}.

Usability is a very important factor in designing an interactive system. If the system is not usable enough for the intended users, it is likely that they do not use the  system so often (underuse) or use the system improperly (misuse) and stick to their current methods for accomplishing the tasks, that both bring costs to the organization or ruin the reputation of the team/company that developed the system \cite{Maguire2001}. There are benefits in designing a usable system both for the developer team and for the customers: increased productivity, reduced errors, reduced training and support, improved acceptance and enhanced reputation \cite{Maguire2001}.
%%---------UCD methods----------------

{
	\renewcommand{\arraystretch}{1.5}
	\renewcommand{\labelitemi}{$\circ$}
	\begin{table}
		
		\scriptsize
		%\small
		\colorbox{gray!20}
		{
			\begin{tabular}{p{0.17\textwidth} p{0.15\textwidth} p{0.17\textwidth} p{0.17\textwidth} p{0.17\textwidth}}
				\hline Planning & Context of use & Requirements & Design & Evaluation\\
				\hline 
				\vspace{-.3cm}
				\begin{flushleft}
					\begin{list}{\labelitemi}{\leftmargin=2pt} 
						\setlength{\itemsep}{0pt}
						\item Usability planning and scoping 
						\item Usability cost benefit analysis 
					\end{list} 
				\end{flushleft}
				&  
				\vspace{-.3cm}
				\begin{flushleft}
					\begin{list}{\labelitemi}{\leftmargin=2pt}
						\setlength{\itemsep}{0pt}
						\item Identify stakeholders
						\item Context of use analysis
						\item Survey of existing users
						\item Field study or user observation
						\item Diary keeping
						\item Task analysis
					\end{list} 
				\end{flushleft}
				& 
				\vspace{-.3cm}
				\begin{flushleft}
					\begin{list}{\labelitemi}{\leftmargin=2pt}
						\setlength{\itemsep}{0pt}
						\item Stakeholder analysis
						\item User cost-benefit analysis
						\item User requirements interview
						\item Focus groups
						\item Scenarios of use
						\item Personas
						\item Existing system or competitor analysis
						\item Task or function mapping
						\item Allocation of function
						\item User, usability and organizational requirements
					\end{list} 
				\end{flushleft}
				& 
				\vspace{-.3cm}
				\begin{flushleft}
					\begin{list}{\labelitemi}{\leftmargin=2pt}
						\setlength{\itemsep}{0pt}
						\item Brainstorming
						\item Parallel design
						\item Design guidelines and standards
						\item Storyboarding
						\item Affinity diagram
						\item Card sorting
						\item Paper prototyping
						\item Software prototyping
						\item Wizard-of-Oz prototyping
						\item Organizational prototyping
					\end{list} 
				\end{flushleft}
				& 
				\vspace{-.3cm}
				\begin{flushleft}
					\begin{list}{\labelitemi}{\leftmargin=2pt}
						\setlength{\itemsep}{0pt}
						\item Participatory evaluation 
						\item Assisted evaluation 
						\item Heuristic or expert evaluation 
						\item Controlled user testing
						\item Satisfaction questionnaires
						\item Assessing cognitive workload  
						\item Critical incidents 
						\item Post-experience interviews 
					\end{list}
				\end{flushleft}
				\\  
				\hline
			\end{tabular} 
		}
		\caption{Methods for user-centered (human-centered) design taken from \cite{Maguire2001}}
		\label{tbl:ucd-methods}
	\end{table}
}
%-------------------------------------

To develop a usable interactive system, both functional requirements and nonfunctional requirements, including usability requirements, should be considered. Traditionally, the focus of software design processes have been on functional requirements, but nowadays there are design frameworks integrating these two \cite{Maguire2001}. According to Shneiderman \cite{Sharp2007}, usability requirements are of five types: learnability, efficiency, memorability, errors, and satisfaction.

By involving users in the design and development process of a system, the system will be more usable for the intended users \cite{Vredenburg2002, ISO13407, IBM2009}. The design approach which places emphasis on involving users in the design is called user-centered design (UCD) or human-centered design (HCD) \cite{Vredenburg2002,ISO13407}.
The main focuses of UCD are active user involvement in the design process, multidisciplinary design teams and iterative design \cite{Maguire2001}. UCD is not a substitute for software development methods, but a complementary process to them.
The UCD process is depicted in Figure~\ref{fig:ucd}. The process starts with planning then context of use analysis, requirements specification, design and evaluation are repeated until the user is satisfied with the design and usability requirements are achieved.
%There are number of principles that are recommended to be considered in UCD \cite{Gulliksen2003}: multidisciplinary design team, understanding users and context, active user participation, early prototyping, continuous evaluation, and holistic design.

%%---------UCD methods----------------
Various methods are defined to support UCD. A broad range of methods are specified by Maguire \cite{Maguire2001}. 
This collection is considered to be a proper introduction of various well-known methods and their relation to different UCD steps.  The methods are summarized in Table~\ref{tbl:ucd-methods}, column names correspond to different steps in the UCD process.

%\section{Previous Works}
%\label{sec:back}
%--------------------------------------------------------------------
%-------------------------------------------------------------------------
%-------------------------------------------------------------------------
\chapter{The Research Process}
\label{ch:process}

Oates \cite{Oates2006} describes a \emph{model of the research process} which consists of various categories of methods such as research strategy, data generation methods, and data analysis. 

Oates describes how self-experiences and motivations along with the knowledge gained from reviewing the literature in the field and being informed about the existing gaps lead to a  ``conceptual framework'' and ``research questions'' (see Section~\ref{sec:qs}).

A conceptual framework clarifies a researcher's way of thinking and structuring a research topic and the research process undertaken \cite{Oates2006}. A conceptual framework includes the research topic and its comprising factors, the \emph{research methodology} (a combination of strategies and methods used), the data analysis approaches, the development methodology, and the research evaluation approach.
Details about the conceptual framework in this study is presented in Section~\ref{sec:cons-frame}.

In order to conduct a research study, a research strategy should be selected (see Section~\ref{sec:strategy}), also data generation methods should be applied in order to gather data and finally analyze the data using qualitative or quantitative methods. In this study, data generation methods are those methods applied in the system development process (see Section~\ref{sec:ucd}) along with the literature published on the topic of interest, i.e. CDS (referred to as ``documents'' in the Oates's model).
Both qualitative (used as part of the system development process) and quantitative  data analysis methods are applied in this study in order to analyze data.
Oates's model is depicted in Figure~\ref{fig:research-process}. This figure also highlights how the process of this study fits into this model.

%-------------------------------------------------------------------
\begin{figure}
	\begin{center}
		\includegraphics[angle=90,height=0.9\textheight]{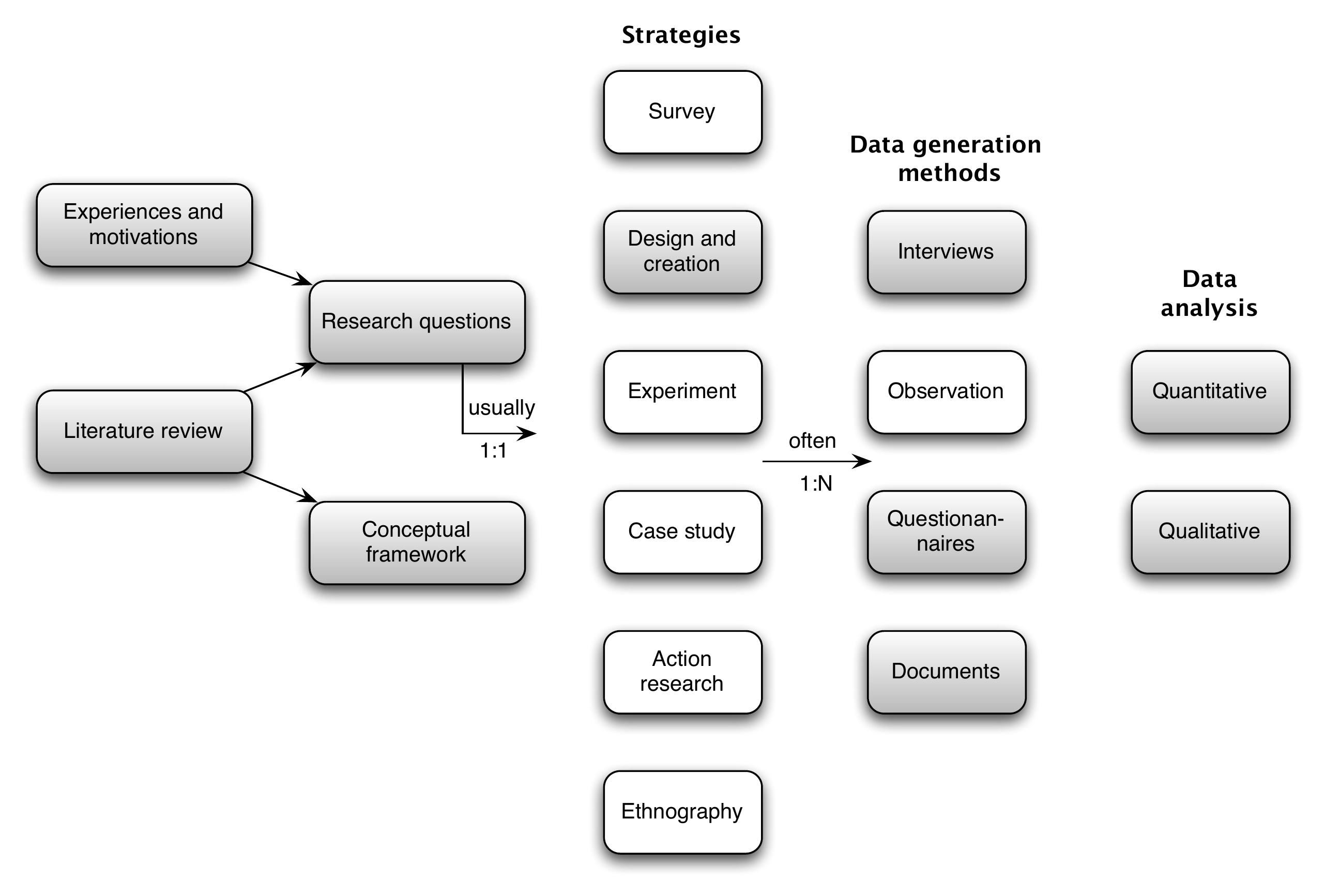}
		\caption{Model of research process adapted from \cite{Oates2006}. The methods applied in this study are colored in gray.}
		\label{fig:research-process}
	\end{center}
\end{figure}
%-------------------------------------------------------------------
%-------------------------------------------------------------------
\section{Conceptual framework}
\label{sec:cons-frame}
As mentioned previously, there have been various efforts in defining CDS success factors and development difficulties ranging from a local, and practical view \cite{Osheroff2005, Arsand2008, Horasani2003, Sittig2008}  to wider national-level views that have tried to provide guidelines or road-maps for developing more effective CDS \cite{Kawamoto2007, Cho2010, Osheroff2007, Huang2007}.

Accordingly, four main categories of challenges and success factors faced in design and development of CDS can be defined:
\begin{itemize}
	\item \textbf{technical issues} concerning knowledge representation and reasoning and also  maintenance of  knowledge
	\item \textbf{integration into EHRs} that deals with the integration to the underlying IT framework or more specifically EHR systems in the clinical organizations which is a sub class of the ``technical issues'', but because of its importance, is considered as a separate category in this thesis.
	\item \textbf{human-computer interaction issues} that focus on the user interface design of these systems and the way users interact with them. User satisfaction, effectiveness and acceptability of CDS  in  a practical setting, and involving clinicians in the design and development of CDS all belong to this category.
	\item \textbf{cultural and organizational issues} that deal with the higher level aspects of motivations, utilizations, monitoring and acceptance of these systems at local, national and international levels.
\end{itemize}

According to the road-map for the United States national action on clinical decision support \cite{Osheroff2007}  to reach widespread adoption of effective CDS, it is crucial that system developers be supported to design ``easy to deploy and use'' applications. It is also recommended that best practices in system development should be disseminated so that other developers can learn from previous successful experiences.

As discussed before, the ultimate goal of efforts in the area of medical informatics is to improve the quality of care, specially by introducing and applying EHRs and CDS.  
To improve the quality of health care, neither of these two concepts is an optimal solution individually.
It is shown that to improve quality of health care, EHR systems should be supported by other services such as CDS. 
On the other hand, in order to develop effective CDS and to broaden its adoption, CDS should be integrated into the existing EHR platform in the clinical organizations and HCI should seriously be taken into consideration in designing and developing them.

%Nevertheless, existing theories in both areas have not been applied optimally in the field of CDS.
%More effort should be put into this field in order to fill the gaps in the both areas which is the aim of the research reported in this thesis.
So far, this study has been done in relation to two of the categories of challenges and success factors in developing CDS namely integration of CDS to EHRs and taking HCI into account in designing and developing CDS. 
Figure~\ref{fig:framework} depicts the factors the comprise this study.
%What do we mean by "Towards usable open-EHR aware clinical decision support?"
In the following, the research questions (Section~\ref{sec:qs}) and objectives (Section~\ref{sec:obj}) are presented.
Chapter~\ref{ch:methods} includes more information about the research strategy and methods applied to answer the research questions.
%-------------------------------------------------------------------
\begin{figure}[!ht]
	\begin{center}
		\includegraphics[scale=0.5]{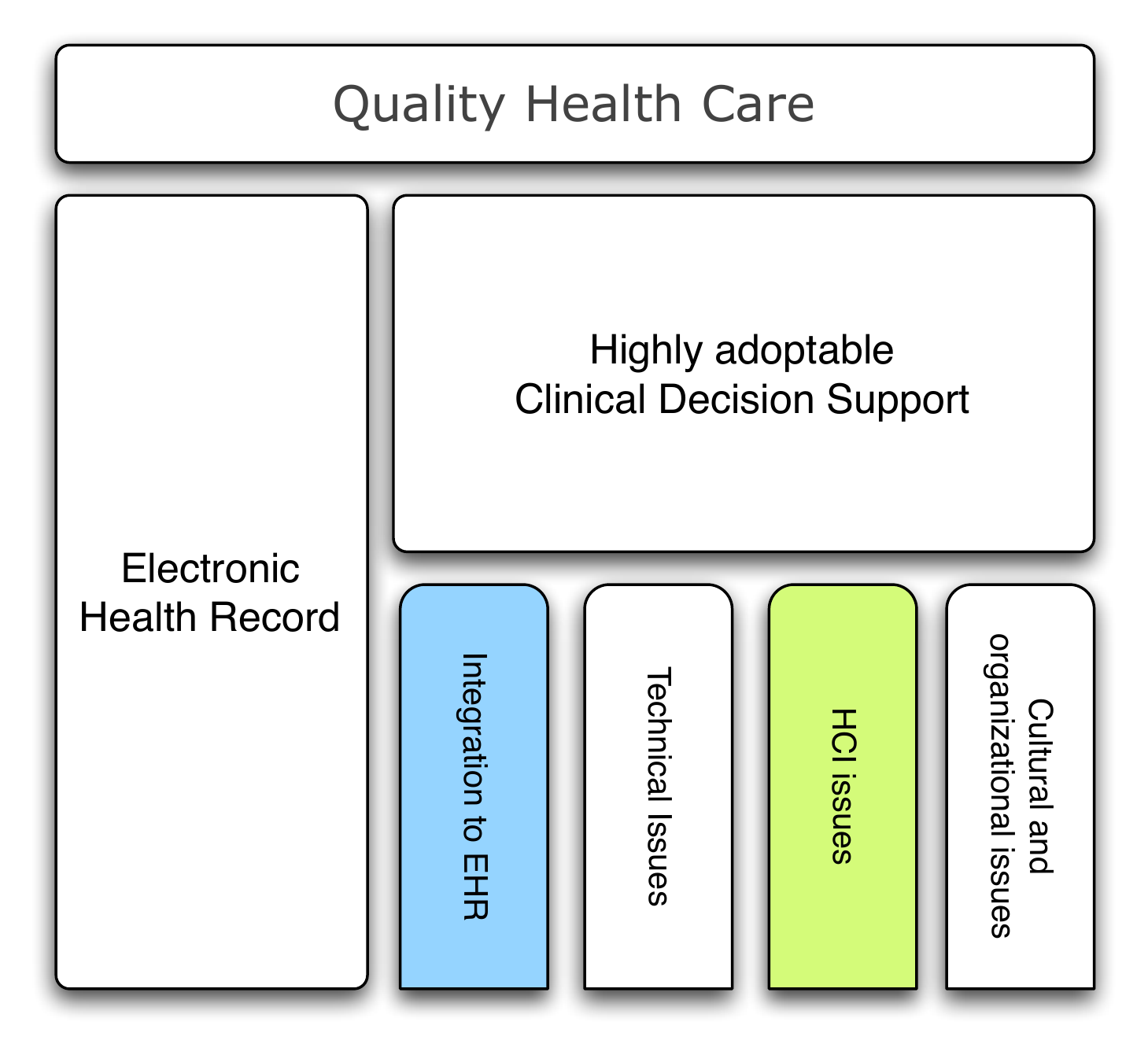}
		\caption{In order to improve the quality of health-care focus on two areas are inevitable: adopting clinical decision support and adopting electronic health records (EHR).
			Development of highly acceptable clinical decision supports is dependent on its integration into EHRs, and also consideration of human-computer interaction (HCI) with the aim of developing usable clinical decision support systems (CDSS).
			There are however other issues such as technical considerations (e.g. knowledge representation and reasoning) and cultural and organizational aspects of adopting clinical decision support in health organizations. Among these pillars, technical issues have gained the most attention so far, but interest in other aspects has also been increasing recently.
			This thesis is invovles two aspects (i.e. colored pillars) namely integration of clinical decision support system into EHRs, and taking HCI into consideration with the aim of developing a usable CDSS that is aware of an EHR standard named \op.}
		\label{fig:framework}
	\end{center}
\end{figure}
%-------------------------------------------------------------------
%-------------------------------------------------------------------
\section{Research Questions}
\label{sec:qs}
%-------------------------------------------------------------------------

The aim of this study has been to answer this research question:

\begin{quote}
	How can usable \op-aware clinical decision support be designed and developed in order to improve the quality of health care?
\end{quote} 

In order to answer this question, several sub-problems were set to be investigated:

\begin{description}
	
	\item[RQ1] Are usability of clinical decision support and methods to reach and assure usability taken into consideration by developers of clinical decision support?
	
	\item[RQ2] Are integration of clinical decision support into electronic health records and adopting electronic health record standards taken into consideration by developers of clinical decision support?
	%empirically in the literature?
	
	\item[RQ3] Is the \op{} suggested approach suitable for designing and developing \op-aware clinical applications, including clinical decision support systems?
	\begin{itemize}
		\item Does the two-level software engineering approach suggested by \op{} work in practice?
	\end{itemize}
	
	\item[RQ4] How can current successful design and development processes such as user-centered design be customized for designing and developing  clinical applications and clinical decision support?
	
	The question involves the following sub-problem:
	\begin{itemize}
		\item  How can the design and development process of an \op-aware clinical application, including clinical decision support systems, be structured with focus on human-computer interaction and involving clinicians in the process?
	\end{itemize}
	
	\item[RQ5] Does \op{} offer any new opportunities for clinical decision support in terms of knowledge representation and reasoning?
	
	The question involves the following sub-problems:
	\begin{itemize}
		\item Can \op{} be used to improve the process of knowledge representation and reasoning in clinical decision support? 
		\item Can clinical decision support benefit from structured, quality validated \op-based electronic health records?
		\item  Is it feasible and practical to integrate clinical decision support interventions into \op-based electronic health records?
	\end{itemize}
\end{description}

To answer the above research questions\footnote{It should be mentioned that the work presented in this thesis is actually related to RQ1-RQ4. RQ5 is suggested as the future direction of this study.}, several objectives should be accomplished. These objectives are defined in the following section.
%-------------------------------------------------------------------
\section{Objectives}
\label{sec:obj}
To investigate \op{} and also UCD in a clinical context with the aim of  answering the research questions, it was decided to develop a CDSS for an oral disease. It was also planned to accomplish the following objectives:

\begin{description}
	
	\item [O1] Literature reviews should be conducted in order to analyze \emph{state of the art} in interplay between HCI and CDS, also the intersection of EHR and CDS.
	
	\item[O2] The \op{} framework should be studied and understood. 
	The \ar{} concept, two-level modeling, and two-level software engineering suggested by \op{} should be analyzed.
	% and its limitation should be specified in order to be improved by the new approach.
	
	\item[O3] A UCD process should be applied from the beginning of the project. 
	Different UCD methods that are applicable for the project should be selected for designing and developing both the CDSS  and \ar s.
	
	\item[O4] Domain-specific information about the disease should be gathered and structured using \op{} \ar s.
	Additionally, as suggested by \op{}, reusable existing \ar s should be specified and customized if applicable.
	
	\item[O5] The strengths and shortcomings of the \op{} approach and the limitations of the two-level software engineering suggested by \op{} should be identified in order to be reconsidered in the proposed approach in this study.
	
	\item[O6] The characteristics of the clinical domain, clinical tasks and clinicians that may have an effect on the user-centered design process should be identified.
	%\item The effect of \ar{} data on GUI should be investigated.
\end{description}
%-------------------------------------------------------------------
%-------------------------
%-------------------------------------------------------------------------
%-------------------------------------------------------------------------	
\chapter{Methods and Tools}
\label{ch:methods}

In this section, the methods and tools applied in order to answer the research questions (see Section~\ref{sec:qs}) are elaborated.
%In this work, a design-based research methodology has been applied \cite{Baumgartner2003}. 

%--------------------------------------------------------------------------------------
\section{Literature Review}
\label{sec:review}

\emph{Literature review} is a research methodology that aims at summarizing the available literature on a topic and presenting an analysis based on that and providing a full picture on the topic \cite{Aveyard2007}.
In this study, literature review is used to analyze \emph{state of the art} in two different but related topics namely ``interplay between HCI and CDS development'', and ``Intersection of CDS and EHR''.
The search strategies are discussed more in detail in the papers I and Paper II.
%-------------------------------------------------------------------
%-------------------------------------------------------------------
%--------------------------------------------------------------------------------------
%--------------------------------------------------------------------------------------
%--------------------------------------------------------------------------------------
\section{The Research Methodology}
Oates in \cite{Oates2006} defines research methodology in information systems as a combination of ``research strategy'', ``design and development process'' and ``data generation methods''. This is depicted in Figure~\ref{fig:methodologies}.

In this study a \emph{design and creation research strategy} is used as the research strategy and a \emph{user-centered design process} is used as the development process. More information about these methods can be found in the following sections.
%-------------------------------------------------------------------
\begin{figure}
	\begin{center}
		\includegraphics[width=0.95\textwidth]{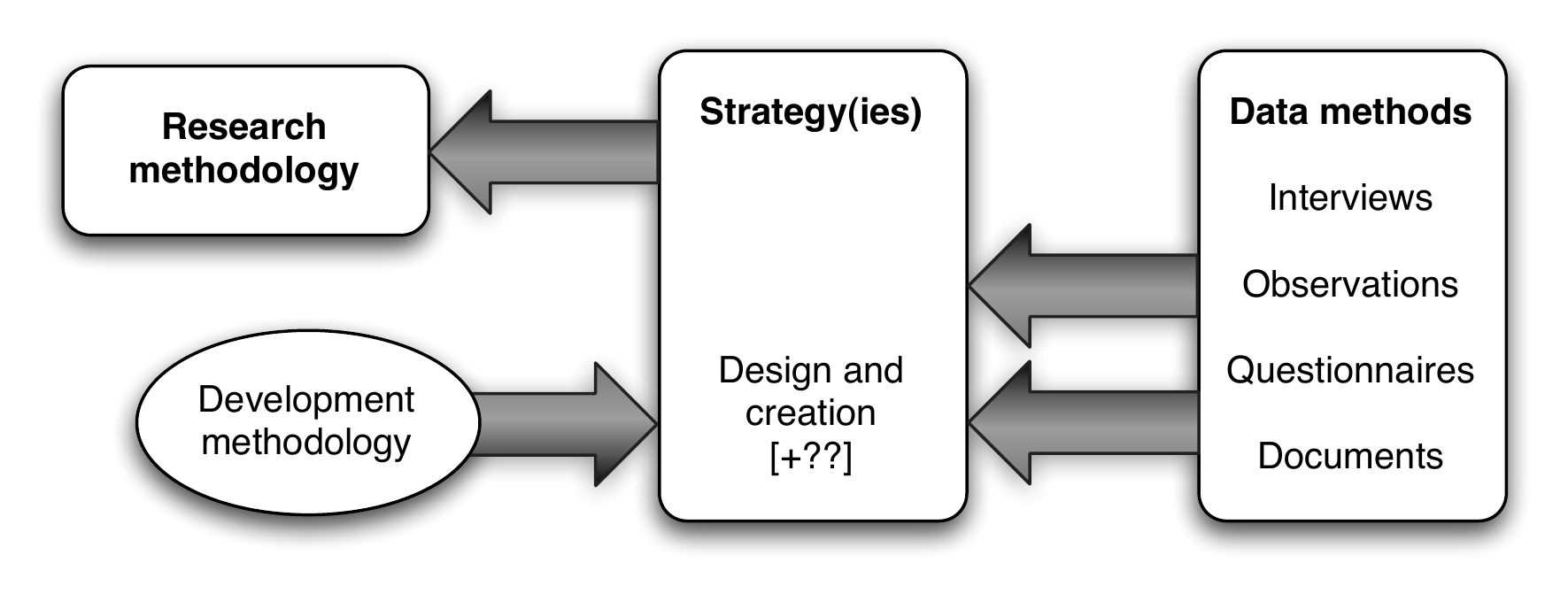}
		\caption{Research methodology and development methodology taken from \cite{Oates2006}. Oates defines research methodology in information systems as a combination of ``research strategy'', ``design and development process'' and ``data generation methods''.}
		\label{fig:methodologies}
	\end{center}
\end{figure}
%-------------------------------------------------------------------
%--------------------------------------------------------------------------------------
\subsection{Design and Creation Research Strategy}
\label{sec:strategy}
The focus of a design and creation research strategy is on developing new software products i.e. artefacts \cite{Oates2006}.
In this study the ``artefact'' to be developed is a \emph{clinical decision support system for \dm}.
To understand this artifact a description of the disease, i.e. \dm, and the characteristics of the CDS are provided in the following.
%---------------------------------------------------------------------------
\subsubsection{A Clinical Decision Support for Dry Mouth}
%We chose to design and develop a CDS for an oral disease named \dm{} in order to investigate \op{} approach and also aspects of applying UCD in a clinical context. 
``Dry mouth or xerostomia is the abnormal reduction of saliva and can be a symptom of certain diseases or be an adverse effect of certain medications'' \cite{Porter2004}. Treatment of Xerostomia is related to finding its cause(s). 
There are five main categories for xerostomia namely drug-induced, disease-induced, radiation-induced, chemotherapy-induced, and cGVHD-induced \cite{Porter2004}. 
Finding cause(s) of \dm{} is a challenge for clinicians and needs to be supported by a clinical application.

A potential \dm{} patient should answer, or a clinician should find an answer to various types of questions such as:
\begin{itemize}
	\item Do you need to moisten your mouth frequently or sip liquids often? 
	\item Have you noticed any swelling of you salivary glands?
	\item Do you smoke or have been smoking regularly?
	\item Are you currently taking 3 drugs or more?
	\item Have you been subjected to therapeutical radiation against your head-and-neck region?
	\item Have you had a feeling of dry mouth daily for more than 3 months?
\end{itemize}

As in other diseases, each of these questions is considered very important in diagnosis. They sound very straightforward, but the difficulty arises when all of these questions should be remembered at the point of care \cite{Walker2006} while the answer provided by a patient should also be supported with an examination by clinicians, e.g. the swollen salivary gland.

The \dm{} CDS is meant to be used in the clinic of oral medicine in Sahlgrenska University Hospital, Gothenburg, Sweden.
Since this system is going to be used integrated with an existing clinical data entry application, i.e. MedView \cite{Jontell2005}, data entry forms are not part of the Graphical User Interface (GUI), however users should be provided with options to edit existing data. Finally, users need to be able to enter their own comments; including diagnosis or treatments to the system. 

The intended decision support process includes these four main steps:
\begin{enumerate}
	\item Presenting an overview of patient-specific information and external knowledge in a way that makes decision making easier
	\item Providing proper reminders and alarms
	\item Helping the user in finding the cause(s) of disease based on the patient's medical record 
	\item Suggesting related materials and treatment options, patient health information and external knowledge
\end{enumerate}

This study has been conducted in oral medicine, however, 
the outcome of the research should be applicable to medical informatics in general, i.e. other diseases, and other clinical applications.
%
%It is worth mentioning that the study started with the aim of developing a CDS that meant to be integrated into an existing clinical application i.e. MedView. However, since the existing system was not \op-based, most of the focus turned to be on dealing with \op-based application development process and creating \ar s related to dry mouth rather than issues specific to CDS such as knowledge representation and reasoning.
%--------------------------------------------------------------------------------------
%--------------------------------------------------------------------------------------
\section{User-Centered Design Process}
\label{sec:ucd}

UCD is a process that places emphasis on involving users in the design \cite{Vredenburg2002,ISO13407}.
%------------------------UCD cycle------------------------------------------------- 
\begin{figure}
	\centering
	\includegraphics[width=0.7\textwidth]{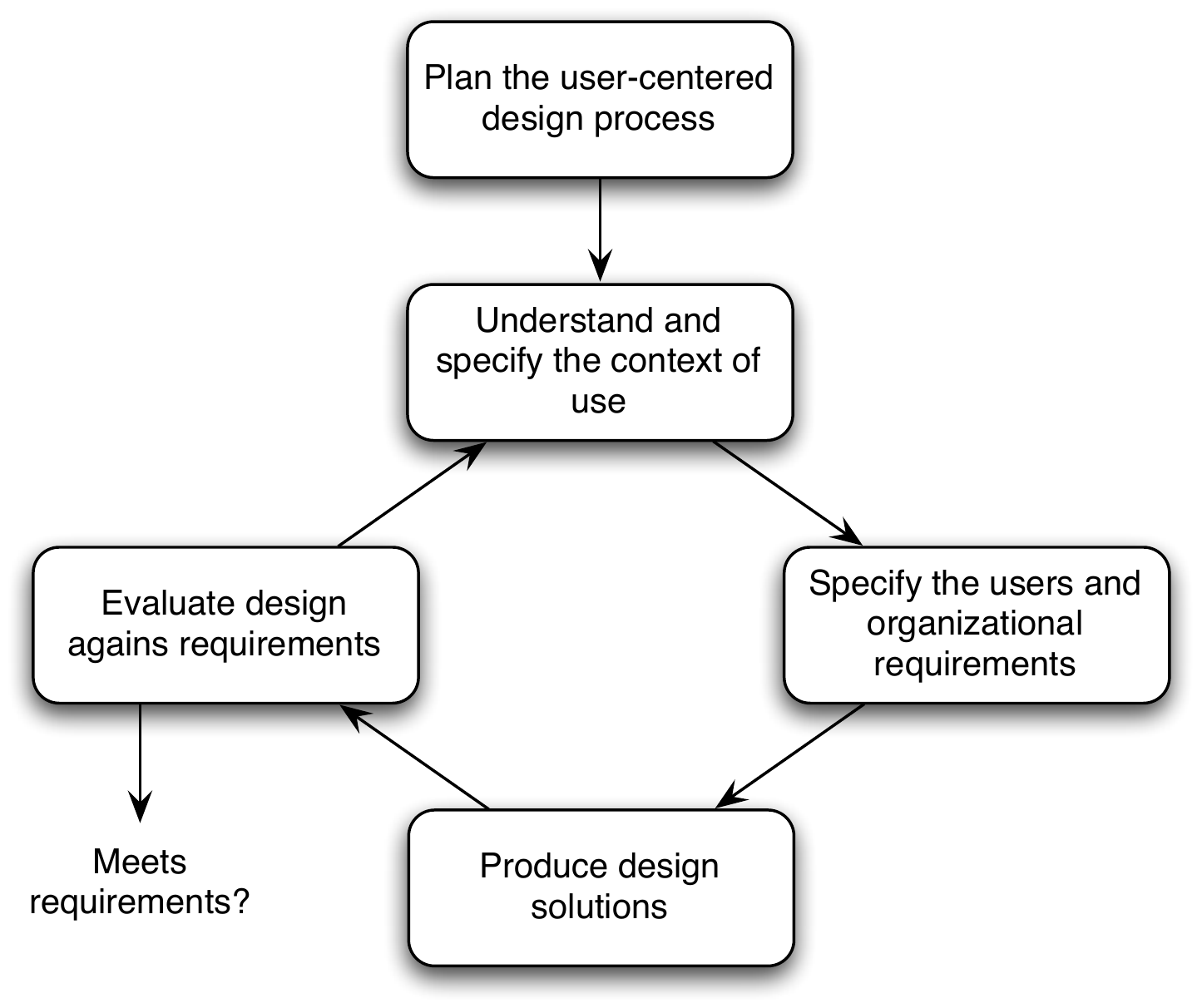}
	\caption{The user centered design cycle \cite{ISO13407, Maguire2001}, see Section~\ref{sec:ucd}}
	\label{fig:ucd}
\end{figure}
%------------------------------------------------------------------------- 
%In this study, UCD has been applied in system design, domain concept modeling and also indirectly in knowledge representation an reasoning.
As depicted in Figure~\ref{fig:ucd}, UCD is a circular design process.
The UCD process consists of five steps \cite{ISO13407}. These steps are 
\begin{enumerate}
	\item plan the human-centered process
	\item understand and specify the context of use
	\item specify the users and organizational requirements
	\item produce design solutions
	\item evaluate design against requirements
\end{enumerate}

This circle will be repeated until the users are satisfied with the design and the requirements are met in the design solution.
%--------------------------------------------
{
	\renewcommand{\arraystretch}{1.5}
	\renewcommand{\labelitemi}{$\circ$}
	\begin{table}
		\centering
		\footnotesize
		%\small
		\colorbox{gray!20}
		{
			\begin{tabular}{p{0.22\textwidth} p{0.22\textwidth} p{0.22\textwidth} p{0.22\textwidth} }
				\hline  Context of Use &  Requirements & Design & Evaluate \\ 
				\hline
				%---
				\vspace{-.3cm}
				\begin{flushleft}
					\begin{list}{\labelitemi}{\leftmargin=2pt}
						\setlength{\itemsep}{0pt}
						\item Identifying stakeholders
						\item Informal context of use analysis
						\item Interviews 
						\item Multidisciplinary group sessions
					\end{list} 
				\end{flushleft}
				%---
				
				&
				\vspace{-.3cm}
				\begin{flushleft}
					\begin{list}{\labelitemi}{\leftmargin=2pt}
						\item Interview
						\item Persona
						\item Existing system or competitor analysis
						\item User, usability and organizational requirements		
						\item Literature study
						\item Domain concept modeling
					\end{list} 
				\end{flushleft}
				
				&
				\vspace{-.3cm}
				\begin{flushleft}
					\begin{list}{\labelitemi}{\leftmargin=2pt}
						\item Brainstorming
						\item Design guidelines and standards
						\item Paper prototyping
						\item Software prototyping
					\end{list} 
				\end{flushleft}
				
				&
				\vspace{-.3cm}
				\begin{flushleft}
					\begin{list}{\labelitemi}{\leftmargin=2pt}
						\item Informal user evluation
						\item Informal expert evaluations
						\item Usability test
						\item Think aloud protocol
						\item Satisfaction questionnaires
						\item Post experience interviews
					\end{list} 
				\end{flushleft}
				
				\\
				%-------
				\hline
			\end{tabular}
		}
		\caption{The UCD methods applied in the project}
		\label{tbl:methods}
	\end{table}
}

As recommended \cite{Gulliksen2003}, a customized UCD process has been applied in this study. The customization was done in order to make the process suitable for the context and also the nature of the project i.e. having \op{} as the underlying EHR standard.
To accomplish the UCD process, several methods were utilized such as prototyping, usability tests, and interviews. Different UCD methods that were applied in this project are summarized in Table~\ref{tbl:methods} and discussed more in detail in Paper V.

The work presented here is the outcome of the first three iterations of this project.
Several users and domain experts were involved in this process.
Several user interface prototypes were developed, evaluated and improved iteratively. The characteristics of this clinical context that have an effect on applying a UCD process were detected and analyzed.
Moreover, UCD was not only applied to reach usability in the design, but also to develop domain concept models to create \ar s. The latter was also accomplished iteratively by involving clinicians (see Paper III).
%-------------------------------------------------------------------------
\subsection*{The Multidisciplinary Project Team}
\label{team}

The project team included the following members:
one specialist in dentistry, who was also one of the stakeholders and initiators of the project (from now on, we refer to this person as the \emph{main clinical partner}), three computer scientists, with knowledge of human-computer interaction, usability and software engineering, and one programmer. 
%------------------------------------------------------------------------- 
\subsection*{Users Involved}

Besides our main clinical partner, who was involved in the project from the beginning, three more specialists in dentistry and one dental hygienist were interviewed during this study both for requirements gathering, \ar s development, and informal evaluation of the user interface paper prototypes (from now on, we refer to this group as \emph{expert panel}). Another group of three specialists in dentistry and one dental hygienist were also asked to participate in the project as test users for user interface evaluations.
%-------------------------------------------------------------------------
%---------------------------------------------------------------------
\section{Assumptions}

The focus of the project is on the design and development process and clinicians' involvement in the process. We assume that \op{} as an interoperability standard is acceptable for our purposes. The aim of this project is ``not'' to prove if \op{} has been successful in EHRs interoperability.

The evaluation process that is mentioned in this thesis refers to the evaluation which is done before releasing the CDS and even before evaluations based on clinical trials which is required to prove reliability of CDS before deployment.

Only the \op{} \ar{} concept is applied in this work and no template is created. 
The idea of \op{} templates is skipped for several reasons: immaturity, no implementation, and finally since it was possible to develop the system without applying them.
%-------------------------------------------------------------------
\section{Archetype Development Process}
\label{sec:arch-dev}
Domain concept modeling (information modeling) was required to understand and specify the data needed to be gathered and presented in the CDS system. Moreover, the domain modeling process was the first step in knowledge gathering for providing CDS.
The \op{} approach suggests \ar s creation as a more structured way of modeling domain concepts.

The \ar{} development was conducted in close collaboration with clinicians, i.e. experts in the domain.
The development process was iterative, this means that domain concept models were created and evaluated by experts in various steps.
More information about the process and tools used to develop \ar s is provided in the following.
%-------------------------------------------------------------------
\subsection*{Iterative Domain Concept Modeling}

The domain concept modeling started with sessions in which our main clinical partner was asked to think about \dm{} and its related concepts and to put as much information as possible on paper. 
% based on the question below 
%\begin{quote}
%What information do you gather about a potential \dm{} patient?
%\end{quote} 
Later, he was asked to prepare a questionnaire based on this question.
The reason for this was that the current clinical system that clinicians in the clinic use in their everyday work is based on the idea of clinical questionnaires.
Questions on the questionnaire were then categorized based on \op{} concepts; in other words, their logical relation, e.g. is it related to the patient's history or examining the patient?
%In the next step, simple diagrams were created based on the questionnaire. 
%More about this process can be found in Section(paper III).
%-------------------------------------------------------------------
\subsection*{Mind Mapping Diagrams}

For better communication of the domain concepts in order to create \ar s, simple diagrams were created based on the questionnaires and also the outcome of the brainstorming sessions.
For this purpose, a mind-map application was used to make it possible for our expert panel to simply understand and edit the created diagrams. 
The mind mapping software used in this step is called \emph{XMind}~\footnote{\url{http://www.xmind.net}}.

%-------------------------------------------------------------------
\subsection*{Evaluation} 

Iterative design of the domain concept model includes evaluations of the current model based on the literature and experts' opinions, and story-based assessment. Information modeling diagrams were improved several times based on the experts' opinions. Several experts were involved in this process to minimize the subjectivity of the design and to be as broad as possible in collecting knowledge. A sample mind map is depicted in Figure~\ref{mind-map}
\begin{figure}
	\centering
	\includegraphics[width=\textwidth]{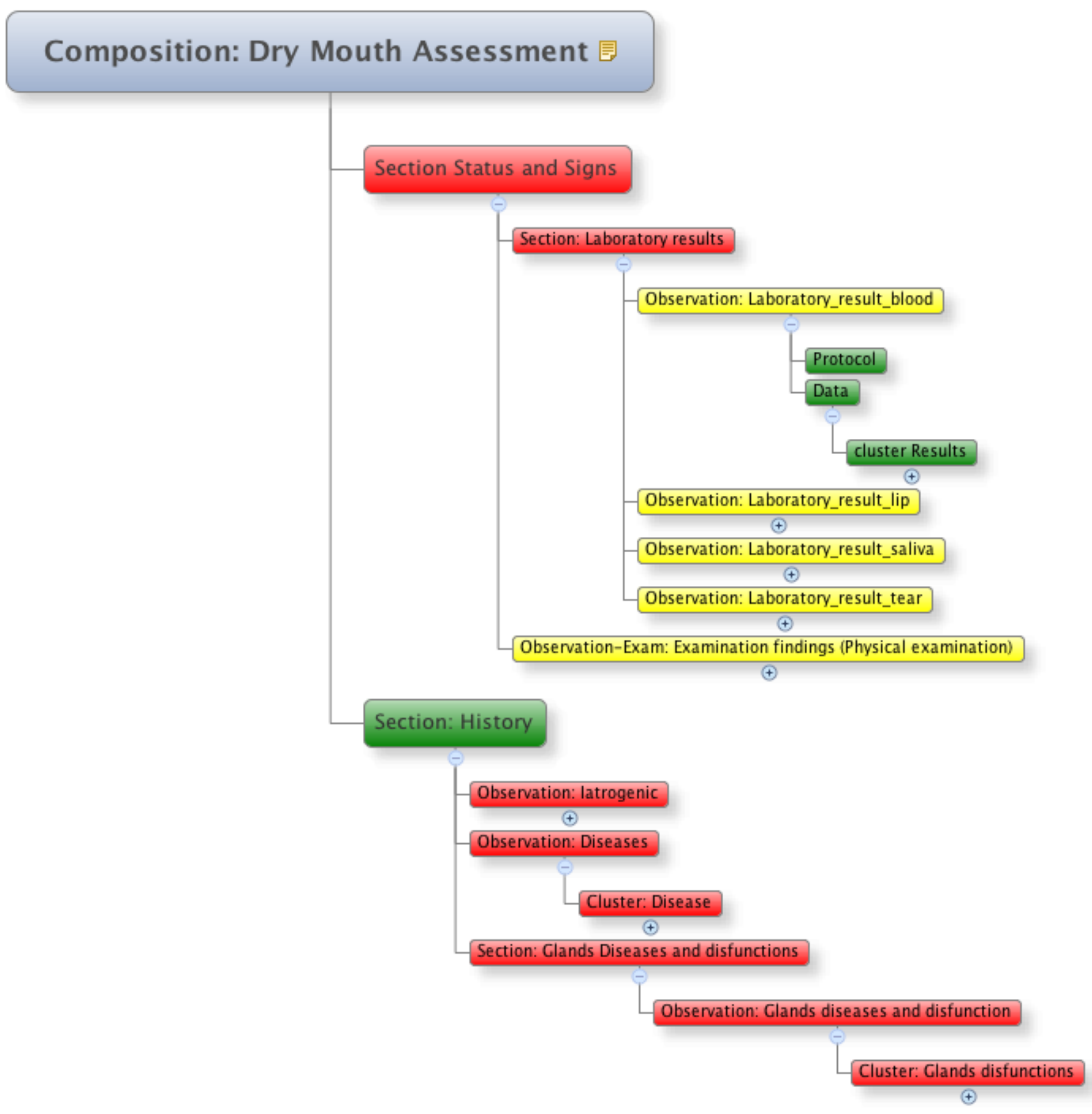}
	\caption{The domain concept modeling (information modeling) was done iteratively and together with the experts in the disease. The XMind application was used in order to easily understand and create diagrams and to communicate information.}
	\label{mind-map}
\end{figure}
%-------------------------------------------------------------------------
\subsection*{Archetype Editor}

For authoring \ar s, a free tool named the \emph{Ocean Archetype Editor}\footnote{\url{http://www.oceaninformatics.com}} was used.
The Ocean Archetype Editor is a visual tool that supports the authoring of \op{} \ar s.
The editor is unicode-enabled, therefore \ar s in any language, including Swedish, can be created in this tool, however, in this project the main language for creating \ar s has been English so far. 
This editor supports full \op{} data types and saves archetypes as different formats such as ADL and XML.
%---------------------------
\begin{figure}
	\centering
	\includegraphics[width=\textwidth]{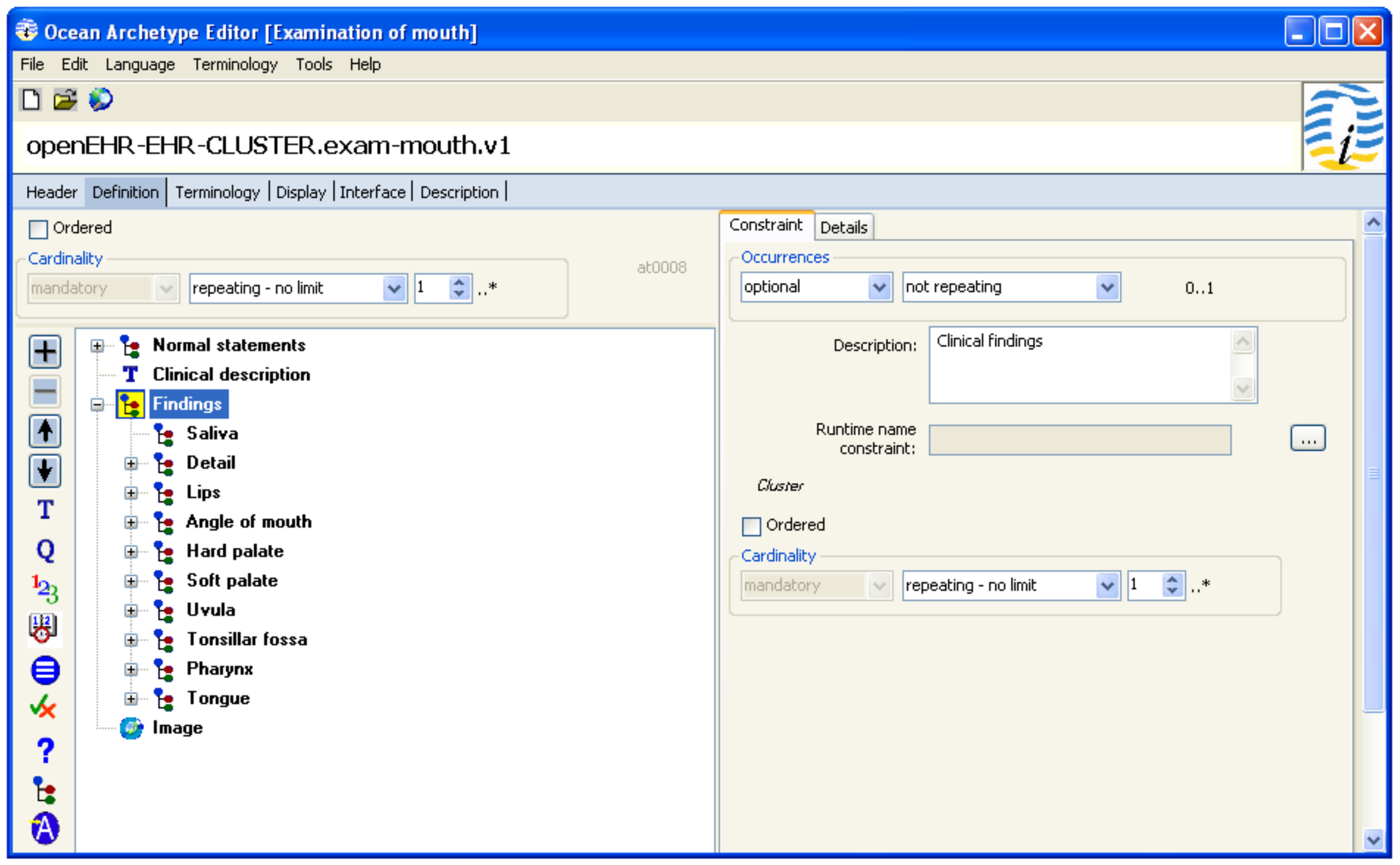}
	\caption{Ocean Archetype Editor is a freely available tool for authoring \op{} \ar s. This tool is meant to be used by clinicians to define specifications of the domain concepts. This tool is developed based on the \op{} reference model and supports various concept types introduced by it.
		In this figure a sample examination \ar{} is being edited. On the left side of the tool, there is a tree structure of the nodes in the \ar. The tools provided several tabs to  support editing various aspects of the \ar{} such as definition and terminology.}
	\label{fig:editor}
\end{figure}
%---------------------------
%-------------------------------------------------------------------------
\subsection*{Reusing Existing Archetypes}
\label{sec:reuse}
It is recommended that whenever possible existing \ar s be reused and/or customized instead of being created from scratch for different local developers.
Accordingly, we have also tried to reuse some of the existing \ar s in this project.
The \op{} community along with other efforts has tried to make the idea of share-ability and reuse-ability of \ar s possible by creating an online repository of reviewed international \ar s. This repository is called The \op{} Clinical Knowledge Manager which is explained below.
%It should be mentioned that our experience showed that the existing published \ar s are not still mature enough. 
%-------------------------------------------------------------------------
\subsection*{The \op{} Clinical Knowledge Manager}

The \op{} clinical knowledge manager (CKM)\footnote{\url{http://www.openehr.org/knowledge}} is an international, online clinical knowledge resource. CKM is a library of clinical knowledge artifacts which at the moment is limited to \op{} \ar s  and templates. It is anticipated that a complementary repository for other related artifacts like terminology subsets be provided in the future. 
This repository provides the foundation for interoperable EHRs. The \op{} \ar s available in the CKM go under a review and publication process in order to be accessible to others.
Users interested in modeling clinical content may participate in the creation and/or enhancement of this international set of \ar s.
%---------------------------
\begin{figure}
	\centering
	\includegraphics[width=\textwidth]{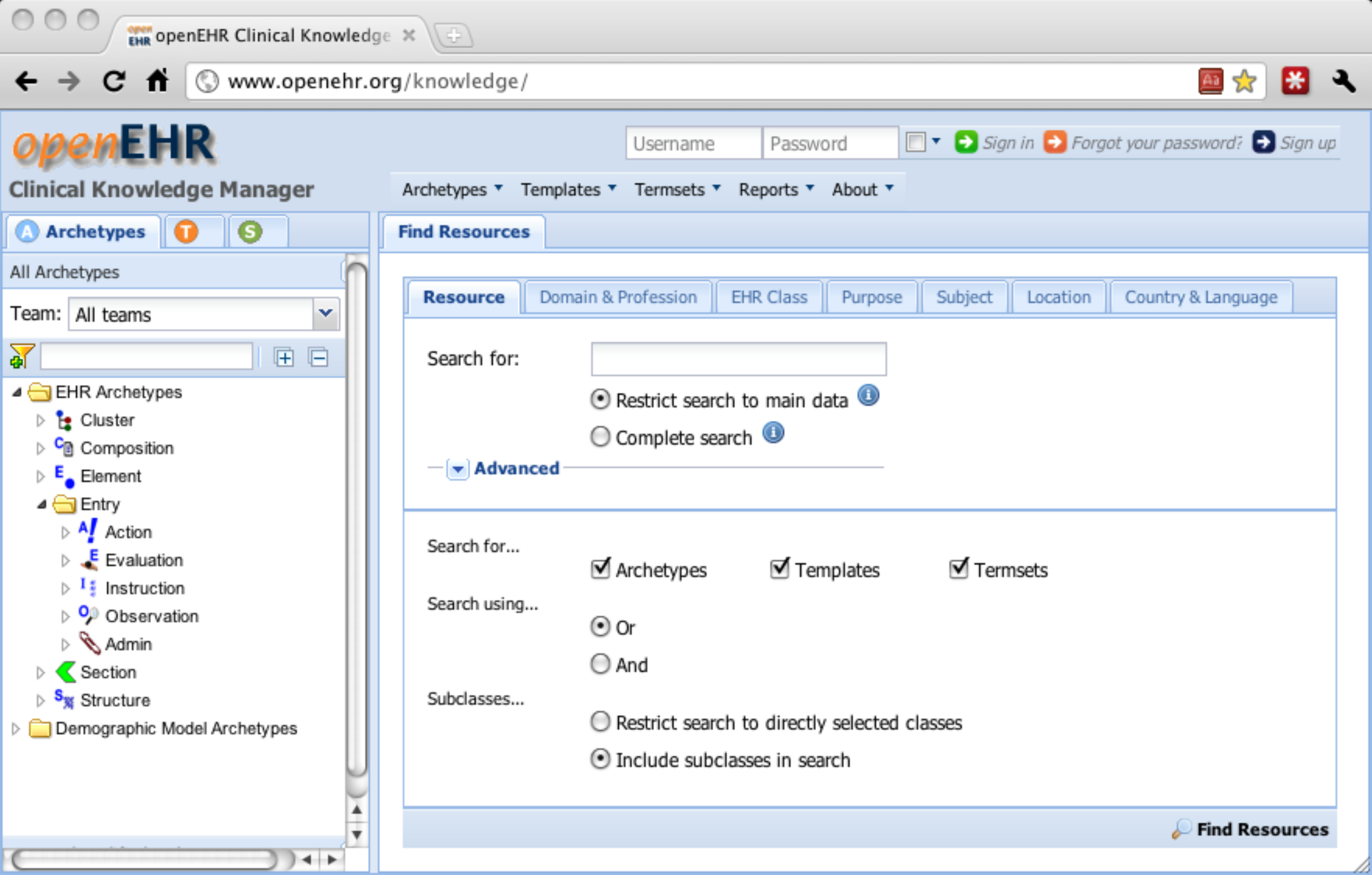}
	\caption{The \op{} clinical knowledge manager (CKM) is a common repository of \ar s. Users interested in modeling clinical content may participate in the creation and/or enhancement of this international set of \ar s via this online repository.}
	\label{fig:CKM}
\end{figure}
%---------------------------
%-------------------------------------------------------------------------
%-------------------------
\chapter{Summary of the Attached Papers}
\label{ch:sum}
%-------------------------------------------------------------------------
In this section, a summary of the appended papers is given, along with how they can be put in relation to each other and the research questions. 
%In the following the contributions are elaborated in more details.
Figure~\ref{fig:papers} depicts different areas covered in the publications.

\section[Paper I]{Paper I}
%{RQ1: Is usability (as a success factor) of clinical decision support and methods to reach and assure usability taken into consideration by developers of clinical decision support? (Paper I)}
\label{sec:lit-hci}

A literature review on interplay between HCI and CDS development is presented in Paper I which is related  to \textbf{RQ1}: \emph{are usability of clinical decision support and methods to reach and assure usability taken into consideration by developers of clinical decision support?} This paper contributes to objective \textbf{O1}.

The paper starts with a brief review of the studies dealing with the question of \emph{which factors should be considered in design and development of CDS to result in an acceptable and effective CDS}, to motivate the importance of HCI, usability and UCD in developing CDSSs.
In order to conduct the literature review, two databases (ScienceDirect\footnote{\url{http://sciencedirect.com}} and PubMed\footnote{\url{http://pubmed.org}}) were searched using boolean combinations of some related keywords (usability, human-computer interaction, user-centered design, clinical decision support, medical decision support). This resulted in a total of 153 studies of which only 17 were relevant to the review.

Various concepts such as iterative design, involving clinicians in design and evaluation, qualitative evaluation methods, usability and UCD were the focus of this review.
More about the findings of this literature review can be found in Section~\ref{sec:rq1}.
%-------------------------------------------------------------------.
\section[Paper II]{Paper II}

Paper II includes a literature review conducted in order to answer \textbf{RQ2}: \emph{Are integration of clinical decision support into electronic health records and adopting electronic health record standards taken into consideration by developers of clinical decision support?} This paper contributes to objectives \textbf{O1} and \textbf{O5}.

The paper motivates the important of integrating CDS into EHRs based on findings by other researchers in the field. It is discussed how CDS and EHRs support each other's  success, and finally improving the quality of care.
Based on searching one database, i.e. ScienceDirect, and using boolean combinations of some related keywords (electronic health record, medical health record, clinical decision support, \op, HL7) a total of 98 studies were found where only 25 of them were relevant to the review.

In addition, since the focus of the thesis has been on \op, a discussion of the causes of low adoption of the \op{} approach is presented in this paper as well.
More about the findings of this literature review and the reasons for low adoption of \op{} can be found in Section~\ref{sec:rq2}.
%-------------------------------------------------------------------
\begin{figure}
	\begin{center}
		\includegraphics[width=0.6\textwidth]{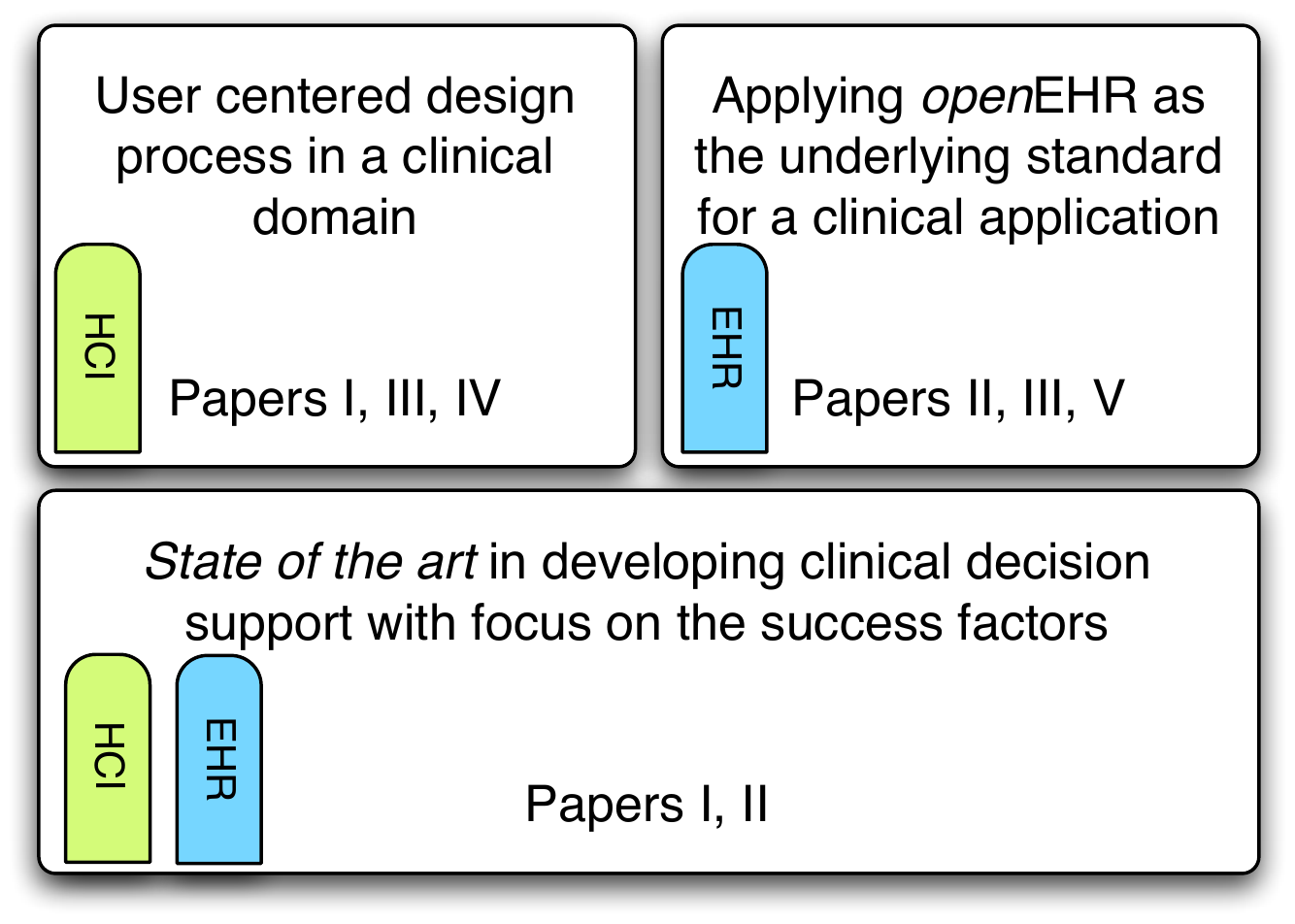}
		\caption{A more empirical view of the study compared to (Figure~\ref{fig:framework}) is presented in this figure. 
			Instead of categorization in an abstract level (i.e. HCI issues,  integration into EHRs) realization of these aspects in form of applying specific approaches (i.e. \op, user-centered design) are introduced here. 
			Moreover, it is shown how these aspects are covered in various publications attached to this thesis.}
		\label{fig:papers}
	\end{center}
\end{figure}
%-------------------------------------------------------------------

%------------------------------------------------------------------
\section[Paper III]{Paper III}

Paper III is related to \textbf{RQ3}: \emph{Is the \op{} suggested approach suitable for designing and developing \op-aware clinical applications, including clinical decision support systems?} and \textbf{RQ4}: \emph{How can current successful design and development processes such as user-centered design be customized for designing and developing clinical applications and clinical decision support?}  This paper contributes to objectives \textbf{O2}, \textbf{O3}, and \textbf{O4}.

This paper describes how a UCD approach can be used in a clinical context for developing an \op-based CDSS. The paper includes a proposed customized UCD approach  along with the preliminary results of designing the GUI, domain concept models and \ar s. Additionally, some challenges faced in adopting \op{} are discussed in Paper III.
%%------------------------------------------------------------------
\section[Paper IV]{Paper IV}

Paper IV is related to \textbf{RQ4}: \emph{How can current successful design and development processes such as user-centered design be customized for designing and developing  clinical applications and clinical decision support?}  This paper contributes to objectives \textbf{O3} and \textbf{O6}.

This paper reports on employing a UCD process in developing a CDSS. Paper IV can be seen as a more detailed version of Paper III, in which the focus has been on the UCD process and the applied methods while details regarding \op{} are skipped in this Paper.
The paper includes results of the three iterations of the project and includes various prototypes of the system, evaluations and analysis of the evaluation results.
In addition, those characteristics of the clinical context that have an effect on applying a UCD process are identified and analyzed in the paper.
%%------------------------------------------------------------------
\section[Paper V]{Paper V}

Paper V is indirectly related to \textbf{RQ3}. By ``indirect'' we mean the paper does not include an answer to this question but is a more practical effort dealing with one of the weaknesses of \op{} that has been discussed in Paper II and Paper III.
In this paper, we have dealt with the question: \emph{how can developers of \op-based clinical applications connect iteratively designed and evaluated user interfaces to the underlying framework with minimum effort?} This paper contributes to objective \textbf{O2} and \textbf{O5}.

In this Paper, a framework for binding pre-designed GUIs to openEHR-based backends is proposed.
The proposed framework contributes to the set of options available for developers.
This approach can be useful especially for various small
scale and experimental systems as well as systems in which the quality of the user
interface is of great importance.
%%------------------------------------------------------------------
\section[Paper VI]{Paper VI}

Paper VI is indirectly related to \textbf{RQ5}. This means that the paper does not cover the answer to the research question but includes discussions about the opportunities \op{} may provide for knowledge representation and reasoning in CDSSs.

In this paper, a software architecture for the CDSS for \dm{} is proposed. The architecture benefits from an existing \op{} framework and also a case-based reasoning (CBR) framework. Case-based reasoning is a knowledge representation and reasoning method that has been popular in the clinical domain and, based on the available domain knowledge and patient data, seems to be a proper choice for this project as well.\\
The paper also includes a methodological approach to developing \op{} \ar s. In addition, motivations for selecting the knowledge representation and reasoning method are given in the paper.
%-------------------------
%------------------------------------------------------------------------- 

\chapter{Thesis Contributions}
\label{ch:cont}
The major contributions of this study are the result of accomplishing the objectives \textbf{O1}-\textbf{O6} (see Chapter~\ref{ch:sum}) and answering the research questions \textbf{RQ1}-\textbf{RQ4}. This is actually documented in the attached papers as results.
A brief summary of the contributions is provided in the following.
%-------------------------------------------------------------------------
\section{The Answer to RQ1}
\label{sec:rq1}
%Are usability of clinical decision support and methods to reach and assure usability taken into consideration by developers of clinical decision support?}

The aim of efforts in the area of CDS is to develop such systems that result in the wider adoption of CDS and accordingly improvement in quality of health care. Various studies have dealt with the question of \emph{which factors should be considered in design and development of CDS to result in an acceptable and effective CDS}. According to these studies, success factors of CDS can be divided in  two main categories of technical and non-technical (i.e. human-related) factors. Most of these human-related factors, are the issues covered by the HCI discipline and related to the concept of usability. 
HCI suggests methods and approaches to address the human-related (i.e. user-related) factors and to assure usability of the applications.

Based on a literature review (see Paper I), one can conclude that while various researchers have so far introduced human-related factors as factors important in the success of CDSSs, HCI is not still a routine practice in this field. 
Only 17 studies were relevant to the literature review whereas just in ScienceDirect more than 100 practical studies on CDSSs are published.
In particular, when it comes to viewing UCD as a life-long process, very few studies can be found that have covered this aspect in developing a CDSS.
It was observed that some of the recommended UCD methods are not applied or rarely are applied in CDS developments. 
Task analysis, usability expert reviews and heuristic evaluation are some of those rarely applied methods.
Finally, there are still cases in which evaluation of the system (our focus is on qualitative evaluations) is only conducted after system deployment. All in all, there is a need for further adoption of HCI (including usability) in this field.

%-------------------------------------------------------------------------
\section{The Answer to RQ2}
\label{sec:rq2}
%: Is integration of clinical decision support into electronic health records, and adopting electronic health record standards  taken into consideration by developers of clinical decision support?}
Taking standards into consideration in any clinical application (and generally any information system) is very important \cite{Osheroff2005}.
Since CDS operates by utilizing both patient/organizational-specific data and clinical knowledge, it is important to take the standards that support each of these areas into account \cite{Osheroff2005}.

Only 25 studies were found in ScienceDirect  to have considered integration of CDS into EHRs (from more than 100 studies that have documented CDS developments). For more information please refer to Paper II. 
We did not find any study that reports on implementation of a CDS by applying \op.
The only study which considers the intersection between \op{} and CDS is \cite{Chen2009} in which the idea of integrating guideline rules into \op{} \ar s is discussed.

The selected articles were reviewed in order to find out whether they consider any of the standards related to CDS (i.e. EHR standards, guideline representation standards, and terminology or vocabulary standards).
It was observed that standardization of guidelines and integration of guidelines into EHR has been discussed in several studies \cite{Chen2009, Barretto2003, Schadow2001}.

The idea of applying standards even for EHR systems is still not mature enough, and it is not surprising to see that researchers rarely consider this in CDS development. For instance, from the 25 studies we reviewed only 6 had considered EHR standardization (all of them applied HL7).

In conclusion, theory supports the benefits offered by integrating CDS into EHRs, still, a great deal of effort should be put into this in order to reach an acceptable level of integration in practice, especially considering standardization aspects of EHR.

Moreover, if we put the the results of the literature review with focus on HCI (see Section~\ref{sec:lit-hci}), it is observable that  there are only a small number of studies that have considered both HCI and EHR integration while developing CDS as depicted in Figure~\ref{fig:hci-ehr}.

%------------------------------------------------------------------- 
\DTLnewdb{hci-ehr}                                                                                    

\DTLnewrow{hci-ehr}
\DTLnewdbentry{hci-ehr}{Name}{HCI consideration}
\DTLnewdbentry{hci-ehr}{Quantity}{8}

\DTLnewrow{hci-ehr}
\DTLnewdbentry{hci-ehr}{Name}{No HCI consideration}
\DTLnewdbentry{hci-ehr}{Quantity}{17}

\begin{figure}[h!]
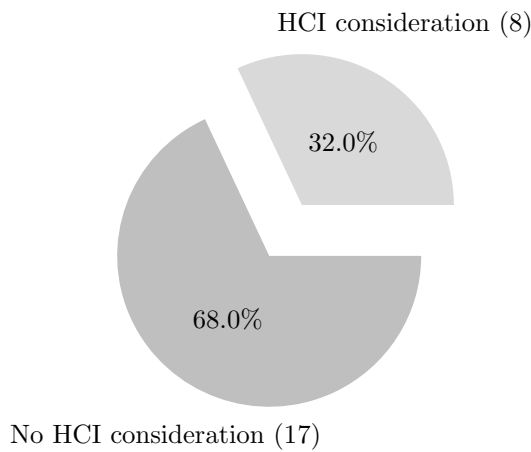

	\centering
	\DTLpiechart{variable=\quantity,innerlabel={\DTLpiepercent\%},outerlabel={\name\ (\quantity)},cutaway={1,2}}{hci-ehr}{\name=Name,\quantity=Quantity}
	\caption{This chart represents how many of the studies have considered integration of CDS into EHRs, as well as HCI in developing CDS.}
	\label{fig:hci-ehr}
	
\end{figure}
%-------------------------------------------------------------------
%-------------------------------------------------------------------------
\section{The Answer to RQ3}
\label{sec:rq3}
%: Is the \op{} suggested approach suitable for designing and developing \op-aware clinical applications, including clinical decision support systems?}
In this study, investigations were carried out into various aspects of developing \op-based applications with the focus on the design and development ``process'' and with the aim of developing ``usable'' CDS.

As discussed in Section~\ref{sec:oph}, \op{} suggests defining various ``roles'' in developing clinical applications, and to divide responsibilities among different roles.
The \op{} two-level software engineering, as might be expected, is compatible with the multi-disciplinary team work suggested in UCD.
The clinicians' expertise can be used by involving them in the domain concept modeling as suggested by \op{} and additionally in user interface design as recommended in the HCI field.
The two-level software engineering (see Figure~\ref{fig:swe}) suggested by the \op{} community is not by itself enough for developing user-friendly applications inasmuch as it does not consider the importance of involving clinicians in designing and evaluating the GUI. 
To develop usable clinical applications, a close collaboration between clinicians and IT developers is needed. Moreover, automatic user interface generation results in interfaces with poor usability. This is  discussed further in the following.

Regardless of its advantages, the \op{} standard suffers from a rather low adoption rate. 
Some possible reasons for the low adoption are the complexity of the standard, lack of documentation and training for developers, and a limited set of tools and frameworks available to ease application development (see Paper II). 
The \op{} community seems to have mostly focused on representing and modeling domain concepts and perfecting the specifications. However, to make \op{} more practical, there is a need for supporting application developers with APIs, frameworks and tools.

Surely, a number of application development projects  exist such as the open source health information platform
\cite{oship-web} (OSHIP), the open EHR-Gen framework \cite{ehr-gen-web}, GastrOS \cite{gastros-web}, and  the \op{} reference framework and application \cite{opereffa-web} (opereffa).
To the best of our knowledge, current \op{} frameworks and tools are based on the idea that clinicians design and create \ar s (and templates)  using existing tools. Later on, a GUI, or some GUI artifacts are generated  based on these \ar s/templates. In order to improve the GUI design, there is a need for manual adjustment of the GUI or its style files (depicted in Figure~\ref{fig:dev-model}-A).
%---------------------  --------------------------------- 
\begin{figure}
	\centering
	\includegraphics[width=0.90\textwidth]{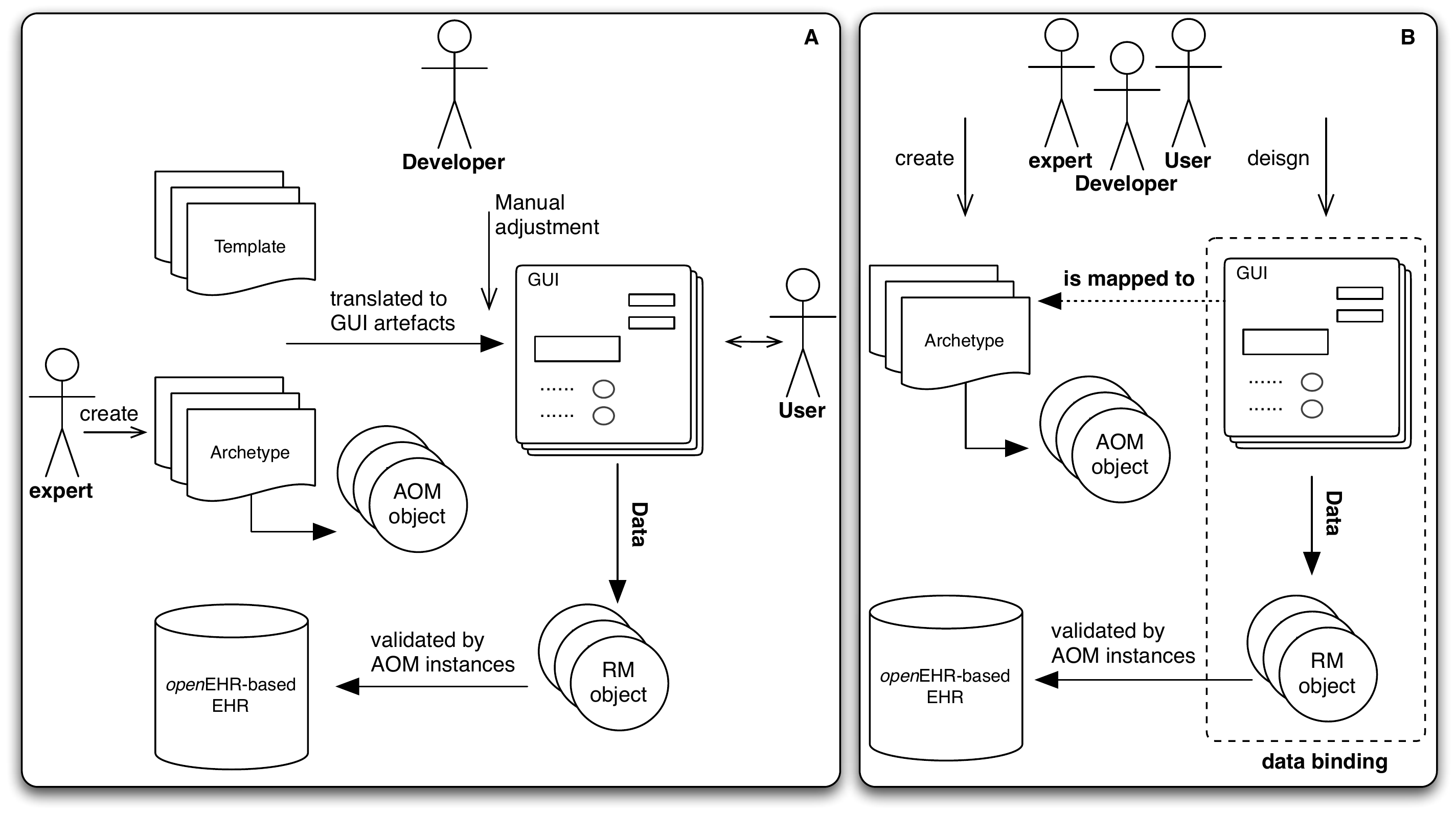}
	\caption{The two development models. The model on the left is supported by \opr}
	\label{fig:dev-model}
\end{figure}
%------------------------------------------------------------------------- 
In contrast to this automatic or semi-automatic approach, there is an alternative approach where there is no generation of GUI based on \ar s. Instead, the interface is designed by experts based on the the users' requirements. Afterwards, there is a need to connect this GUI to the \ar s designed by domain experts (illustrated in Figure~\ref{fig:dev-model}-B).
Unfortunately, the current frameworks do not provide sufficient support for this approach.
Therefore, we have developed an extension, a Java desktop user interface data binding layer, to one of the \op{} application development frameworks, i.e. \opr with the aim of supporting \op{} Java application developers who develop applications according to the aforementioned approach.
%----------CUSTOMIZED UCD--------------------------- 
\begin{figure}[h]
	\centering
	\includegraphics[angle=90, height=0.9\textheight]{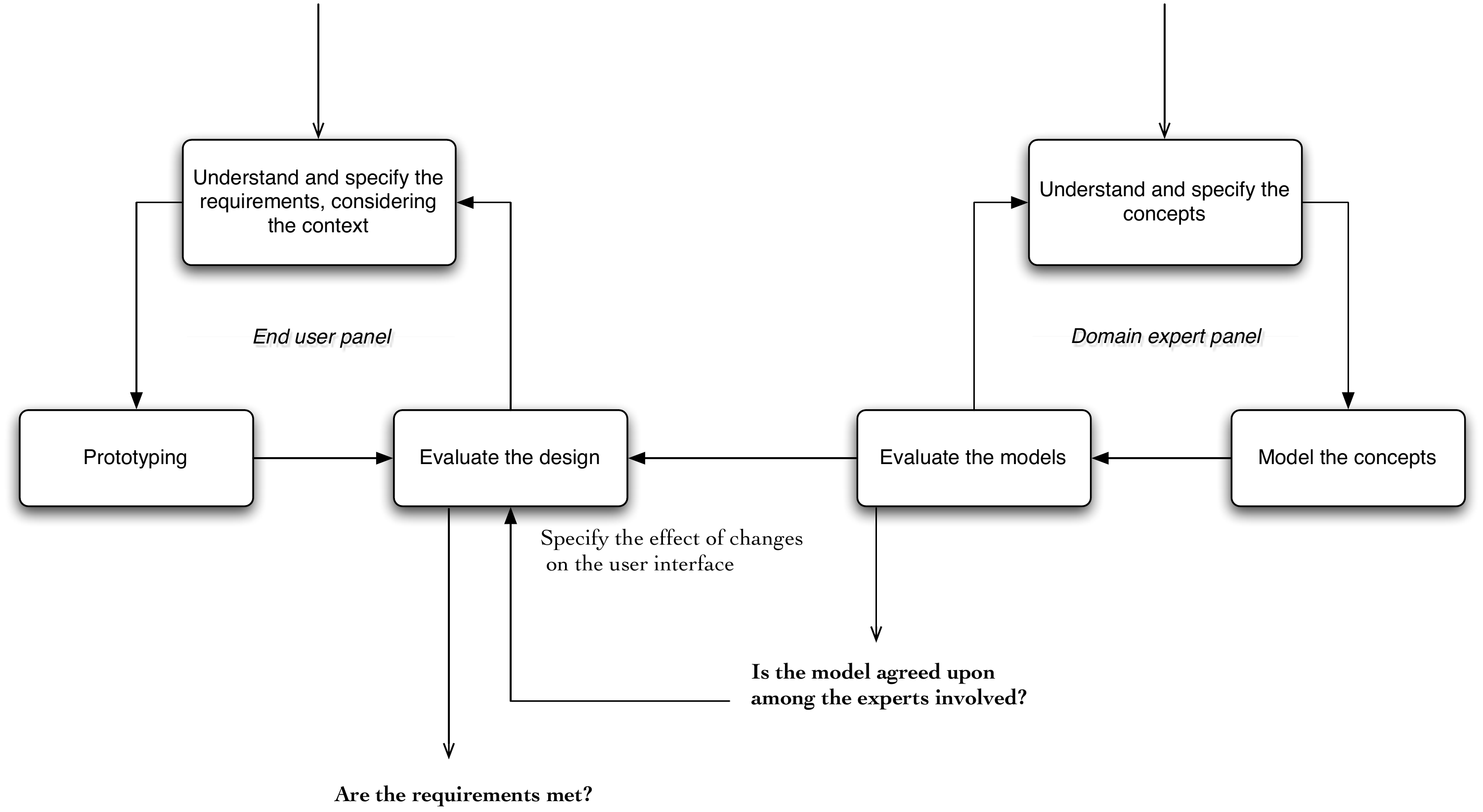}
	\caption{The customized user-centered design applied in this study.}
	\label{fig:cus-ucd}
\end{figure}
%---------------------------------------------------
%-------------------------------------------------------------------------
\section{The Answer to RQ4}
\label{sec:rq4}
%: How can current successful design and development processes such as user-centered design be customized for designing and developing  clinical applications and clinical decision support?}
In this study, we have applied a design and development process that combines UCD and \op{} principles.
The suggested approach considers active involvement of the clinicians in design and evaluation of the \ar s, and also the user interface.
Moreover, the effect of the \ar s on the user interface has been taken into consideration. This customized UCD approach is depicted in Figure~\ref{fig:cus-ucd}. 
The proposed UCD process is compatible with the \op{} software development approach illustrated in Figure~\ref{fig:dev-model}-B (see the previous section for details)
More about this UCD process can be found in Paper III. 

In addition, we have tried to learn from applying UCD in a clinical context. Characteristics of the context, users and tasks that may have an effect on applying UCD are also identified in this study. These characteristics should be taken into consideration in design and development of clinical applications including CDS (see Paper IV).
%------------------------------------------------------------------------- 
\chapter{Future Work}
\label{ch:fut}
%----------------------------------------------------------------------------------------------------------

The main future direction of this study is to address \textbf{RQ5}:
\begin{quote}
	Does \op{} offer any new opportunities for clinical decision support in terms of knowledge representation and reasoning?
\end{quote}

Various sub-problems related to this \textbf{RQ} would be:
\begin{enumerate}
	\item Can \op{} be used to improve the process of knowledge representation and reasoning in clinical decision support?
	\item Can clinical decision support benefit from structured, quality validated \op-based electronic health records?
	\item Is it feasible and practical to integrate clinical decision support interventions into \op-based electronic health records?
\end{enumerate}  

Moreover, there are other aspects of the study that need more investigations:
\begin{itemize}
	\item What are the challenges in applying user-centered design in a clinical context and how to tackle these challenges?
	%\item Does user-evaluations and expert-evaluations of the clinical applications have the same effect in discovering usability problems?
	\item Is the idea of automatic user interface generation acceptable from a human-computer interaction perspective? 
\end{itemize}
%------------------------------------------------------------------------------
\chapter{Conclusion}
\label{ch:conc}

This thesis investigates the question: \emph{how can usable \op-aware clinical decision support be designed and developed in order to improve the quality of health care?} 
In order to answer this question, several sub-problems were identified to be investigated, and accordingly, several objectives to be accomplished. Both theoretical and empirical research strategies were used in order to address the identified research questions (see Chapter~\ref{ch:process}). Of the five specified research questions, four are answered in this thesis.

Analysis of \emph{state of the art} in interplay between HCI and CDS and also the intersection between CDS and EHRs revealed that consideration of both HCI and integration of CDS into EHRs is more appreciated in theory than practice and are yet to be developed (see Paper I and Paper II).

Moreover, the experience in designing an \op-based clinical application revealed that apart from benefits offered by the \op{} approach such as defining different roles and involvement of users in defining domain concepts, there are various shortcomings that should be improved, for instance the insufficient support for \op{} application developers.
%automatic user interface generation offered by \op{} results in user interfaces with poor usability and does not consider the importance of iterative design and evaluation recommended in the HCI domain (see Paper V).
Additionally, it was observed that there are characteristics of the domain, tasks and users in the domain that developers should be informed about while applying UCD methods.

Finally, several future directions of the research were presented with focus on both  the UCD development process, and investigation of \op{} more in depth (see Chapter~\ref{ch:fut}).
%-----------------------------------------------------------------------------
\newpage
%%%%%%%%%%%%%%%%%%%%%%%%%%%%%%%%%%%%%%%%%%%%%%%%%%%%%%%%%%%% References
%\nocite{*}

\bibliographystyle{ieeetr}
\bibliography{ref}
\addcontentsline{toc}{chapter}{Bibliography}

\cleardoublepage
\vspace*{3cm}

\pagenumbering{gobble}

\part{Publications}
\pagenumbering{arabic}
\setcounter{page}{50}

\cleardoublepage
\vspace*{3cm}
\thispagestyle{empty}

\chapter*{Paper I}
\addcontentsline{toc}{chapter}{Paper I}

%%%%%%%%%%%%%%%%%%%%%%%%%%%%%%%%%%%%%%%%%%%%%%%%%%%%%%%%%%%% Paper I
%{\Large\bf\hfill Paper I}
\begin{centering}
\vspace*{30ex}

{\Large \bf Towards Interaction Design \\ in Clinical Decision Support Development: \\A Literature Review \\}

\vspace{0.7cm}
Hajar Kashfi \\  
\vspace{0.5cm}
%To be submitted to:
%\vspace{5ex}
{\it International Journal of Medical Informatics}, Elsevier. \\(manuscript submitted)\\ 

\end{centering}
\cleardoublepage
\vspace*{3cm}
%%%%%%%%%%%%%%%%%%%%%%%%%%%%%%%%%%%%%%%%%%%%%%%%%%%%%%%%%%%% Empty
\thispagestyle{empty}
\newpage
%%%%%%%%%%%%%%%%%%%%%%%%%%%%%%%%%%%%%%%%%%%%%%%%%%%%%%%%%%%% Empty
\includepdf[pages=-]{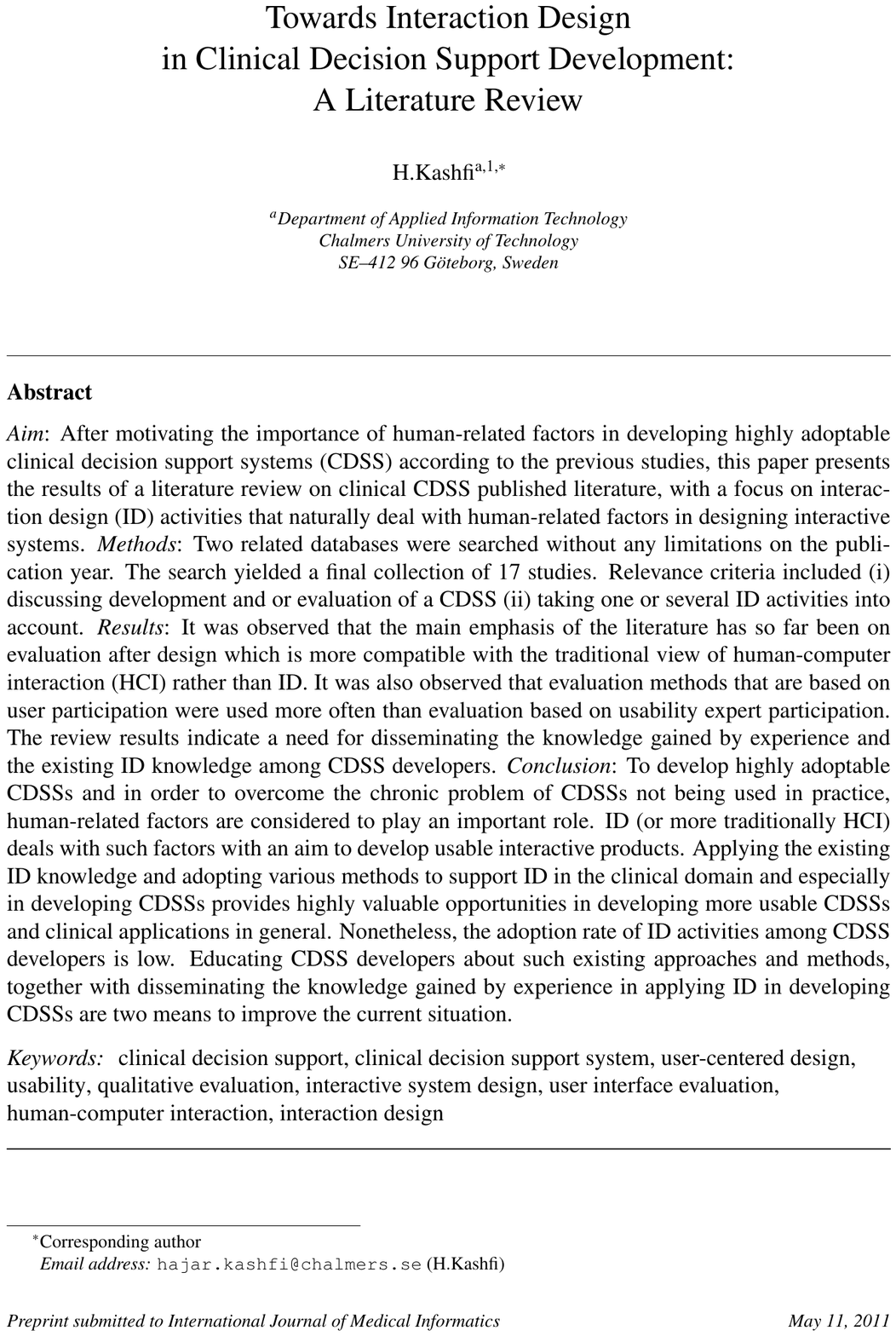}
\cleardoublepage
\vspace*{3cm}
\thispagestyle{empty}

\chapter*{Paper II}
\addcontentsline{toc}{chapter}{Paper II}

%%%%%%%%%%%%%%%%%%%%%%%%%%%%%%%%%%%%%%%%%%%%%%%%%%%%%%%%%%%% Paper II
%{\Large\bf\hfill Paper II}
\begin{centering}
\vspace*{30ex}
{\Large\bf  The Intersection of Clinical Decision \\ Support and Electronic Health Record: \\A Literature Review\\}
\vspace{0.7cm}
Hajar Kashfi \\ 
\vspace{0.5cm}
%\vspace{5ex}
{\it 1$^\mathrm{st}$ International Workshop on Interoperable Healthcare Systems (IHS’2011) - Challenges, Technologies, and Trends}, Szczecin, Poland, September 19-21, 2011.\\  (manuscript submitted)\\ 

\end{centering}
\cleardoublepage
\thispagestyle{empty}
\vspace*{3cm}

%%%%%%%%%%%%%%%%%%%%%%%%%%%%%%%%%%%%%%%%%%%%%%%%%%%%%%%%%%%% Empty
\thispagestyle{empty}
\newpage
%%%%%%%%%%%%%%%%%%%%%%%%%%%%%%%%%%%%%%%%%%%%%%%%%%%%%%%%%%%% Empty
\includepdf[pages=-]{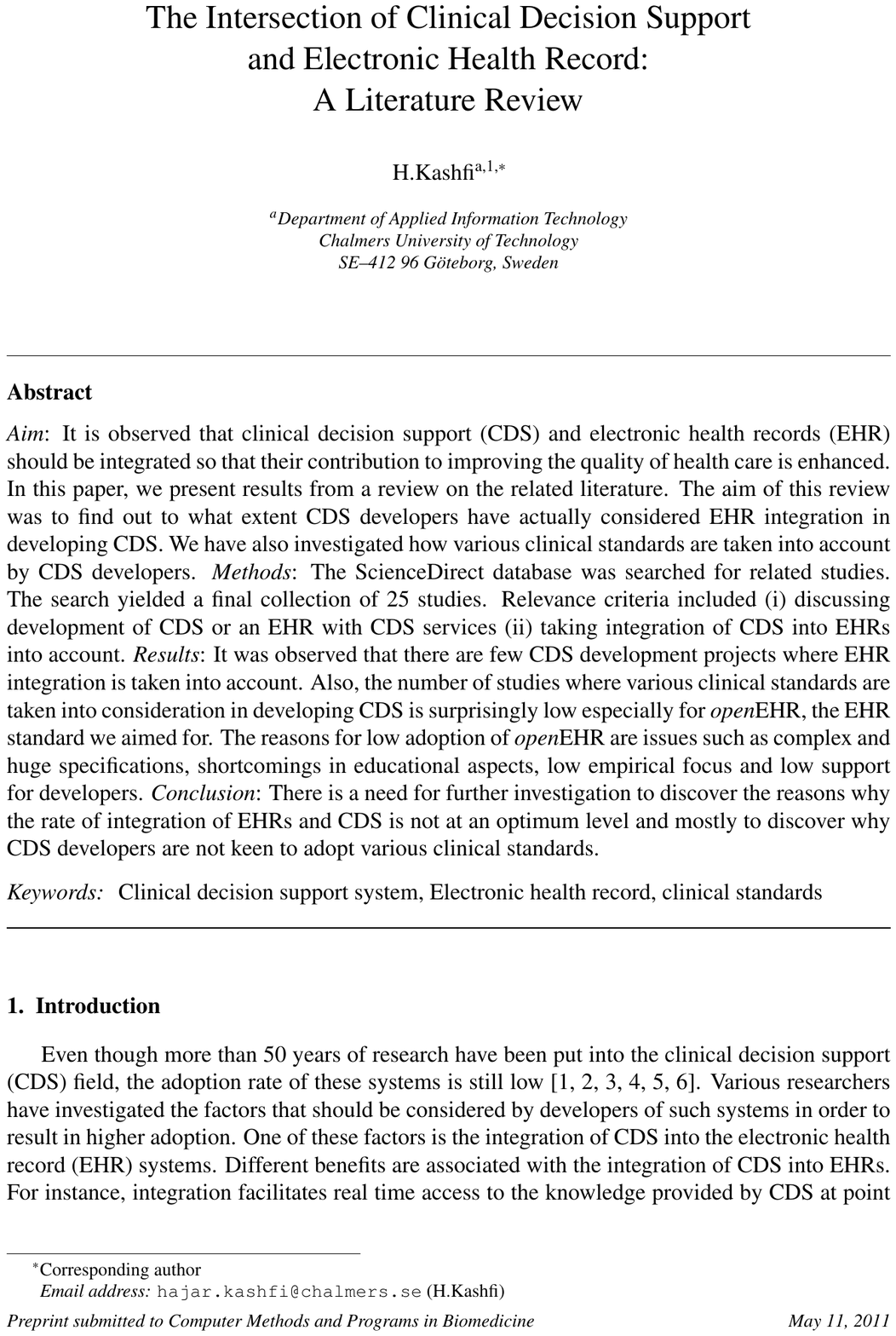}
\thispagestyle{empty}
\cleardoublepage
\vspace*{3cm}

\chapter*{Paper III}
\addcontentsline{toc}{chapter}{Paper III}

%%%%%%%%%%%%%%%%%%%%%%%%%%%%%%%%%%%%%%%%%%%%%%%%%%%%%%%%%%%% Paper III
%{\Large\bf\hfill Paper III}
\begin{center}
\vspace*{30ex}

{\Large \bf Applying a User-centered Design Methodology in a Clinical Context} \\
\vspace{0.7cm}
Hajar Kashfi \\ 
\vspace{0.5cm}
%\vspace{5ex}
{\it The 13$^\mathrm{th}$ International Congress on Medical Informatics (MedInfo2010), Studies in health technology and informatics}, 2010 Jan ;160(Pt 2):927-31. \\ (reprinted with an updated layout)
\end{center}

\cleardoublepage
\thispagestyle{empty}
\vspace*{3cm}
%%%%%%%%%%%%%%%%%%%%%%%%%%%%%%%%%%%%%%%%%%%%%%%%%%%%%%%%%%%% End Paper III
\thispagestyle{empty}
\newpage
%%%%%%%%%%%%%%%%%%%%%%%%%%%%%%%%%%%%%%%%%%%%%%%%%%%%%%%%%%%% Empty
\includepdf[pages=-]{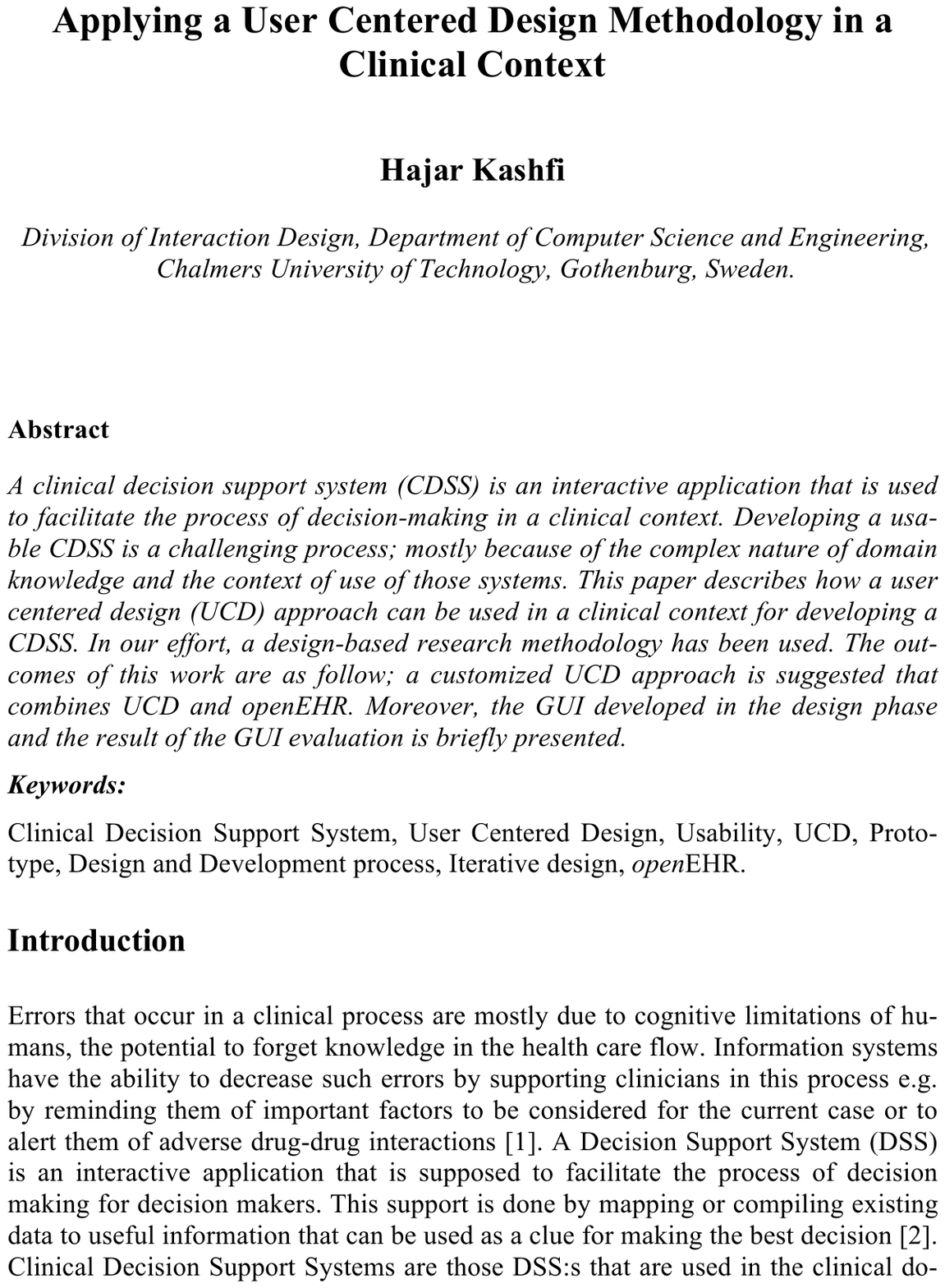}

%\thispagestyle{empty}
%\cleardoublepage
%\vspace*{3cm}

\chapter*{Paper IV}
\addcontentsline{toc}{chapter}{Paper IV}

%{\Large\bf\hfill Paper V}
\begin{centering}
\vspace*{30ex}

{\Large \bf Applying a User-centered Approach in Designing a Clinical Decision Support System} \\
\vspace{0.7cm}
Hajar Kashfi\\ 
\vspace{0.5cm}
%\vspace{5ex}
{\it  Computer Methods and Programs in Biomedicine}, Elsevier.\\ (manuscript submitted)\\ 
\end{centering}
\cleardoublepage
\vspace*{3cm}

\thispagestyle{empty}
\newpage
\includepdf[pages=-]{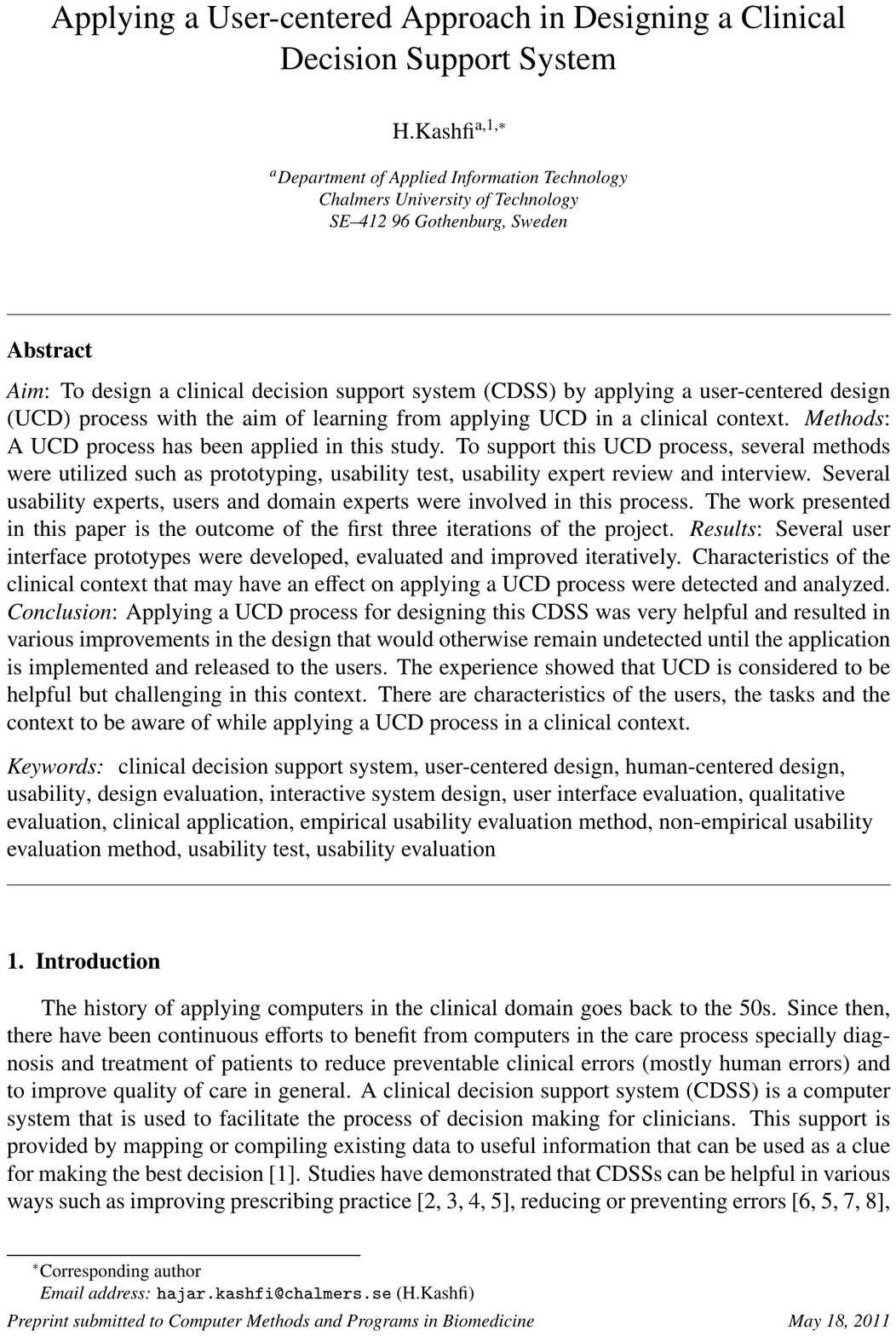}
%%%%%%%%%%%%%%%%%%%%%%%%%%%%%%%%%%%%%%%%%%%%%%%%%%%%%%%%%%%% Empty

%%%%%%%%%%%%%%%%%%%%%%%%%%%%%%%%%%%%%%%%%%%%%%%%%%%%%%%%%%%% Paper V
\thispagestyle{empty}
\cleardoublepage
\vspace*{3cm}

\chapter*{Paper V}
\addcontentsline{toc}{chapter}{Paper V}

%{\Large\bf\hfill Paper V}
\begin{centering}
\vspace*{30ex}

{\Large \bf Supporting \op{} Java Desktop Application Developers} \\
\vspace{0.7cm}
Hajar Kashfi, Olof Torgersson \\
\vspace{0.5cm}
%\vspace{5ex}
%MIE2011
{\it The XXIII  International Conference of the European Federation for Medical Informatics, Proceedings of Medical Informatics in a United and Healthy Europe (MIE2011)}, Oslo, Norway, 28-31 August, 2011.\\ (in print) \\ 
%Oslo, Norway \\
%28-31 August, 2011.\\
\end{centering}
\cleardoublepage
\vspace*{3cm}
%%%%%%%%%%%%%%%%%%%%%%%%%%%%%%%%%%%%%%%%%%%%%%%%%%%%%%%%%%%% End Paper V
\thispagestyle{empty}
\newpage
\includepdf[pages=-]{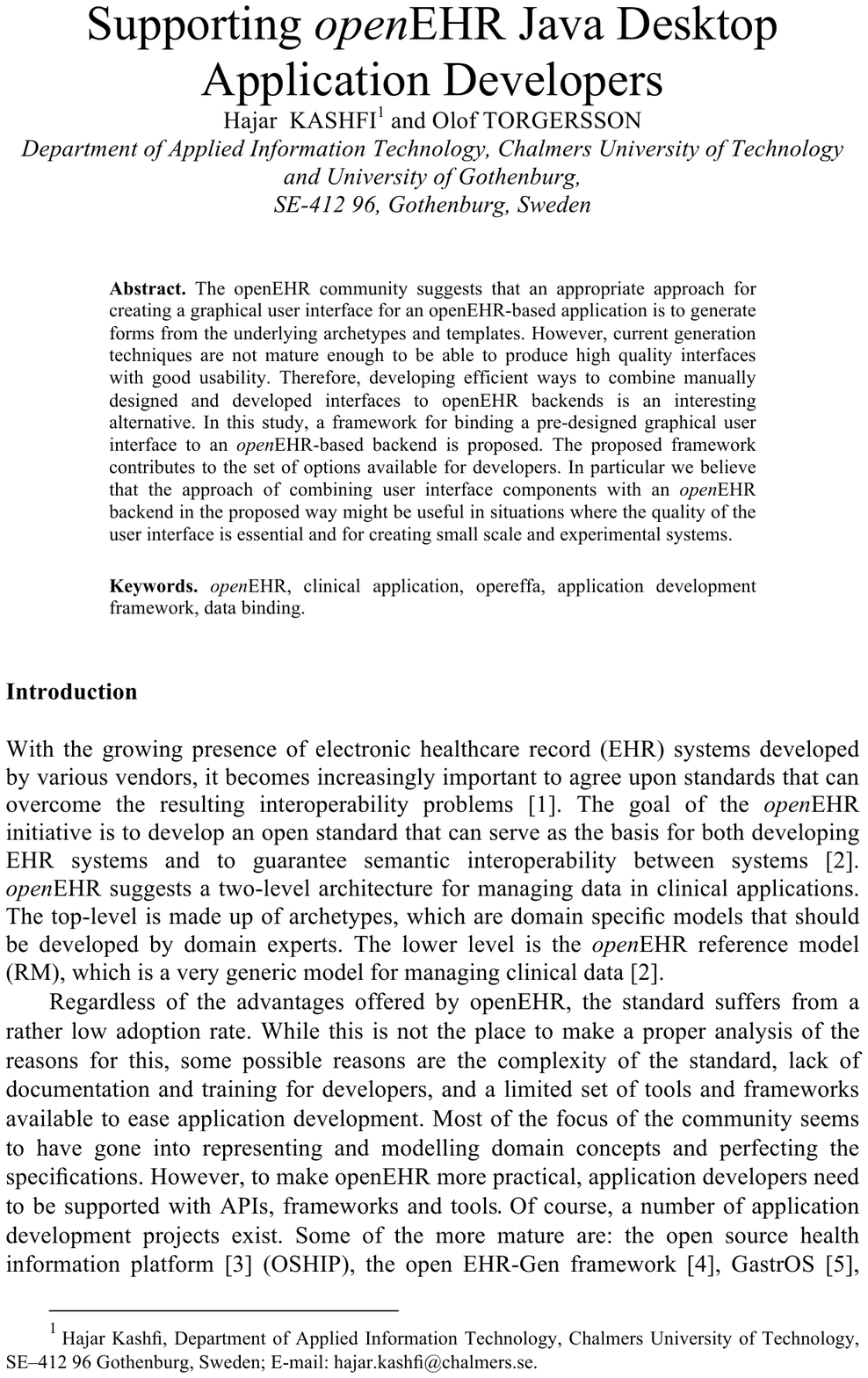}
%%%%%%%%%%%%%%%%%%%%%%%%%%%%%%%%%%%%%%%%%%%%%%%%%%%%%%%%%%%% 
\thispagestyle{empty}
\cleardoublepage
\vspace*{3cm}

\chapter*{Paper VI}
\addcontentsline{toc}{chapter}{Paper VI}

%{\Large\bf\hfill Paper VI}
\begin{centering}
\vspace*{30ex}
 
{\Large \bf Towards a Case-Based Reasoning Method for \\ \op-Based Clinical Decision Support } \\
\vspace{0.7cm}
Hajar Kashfi,   Jairo Robledo Jr.\\ 
\vspace{0.5cm}
%\vspace{5ex}
%KR4HC'11 \\
{\it Proceedings of The 3$^\mathrm{rd}$ International Workshop on Knowledge Representation for Health Care (KR4HC'11)}, Bled, Slovenia, 6  July, 2011.\\ (in print) \\ 
%Bled, Slovenia\\
%6  July, 2011.\\
\end{centering}
\cleardoublepage
\vspace*{3cm}
%%%%%%%%%%%%%%%%%%%%%%%%%%%%%%%%%%%%%%%%%%%%%%%%%%%%%%%%%%%% End Paper V
\thispagestyle{empty}
\newpage
\includepdf[pages=-]{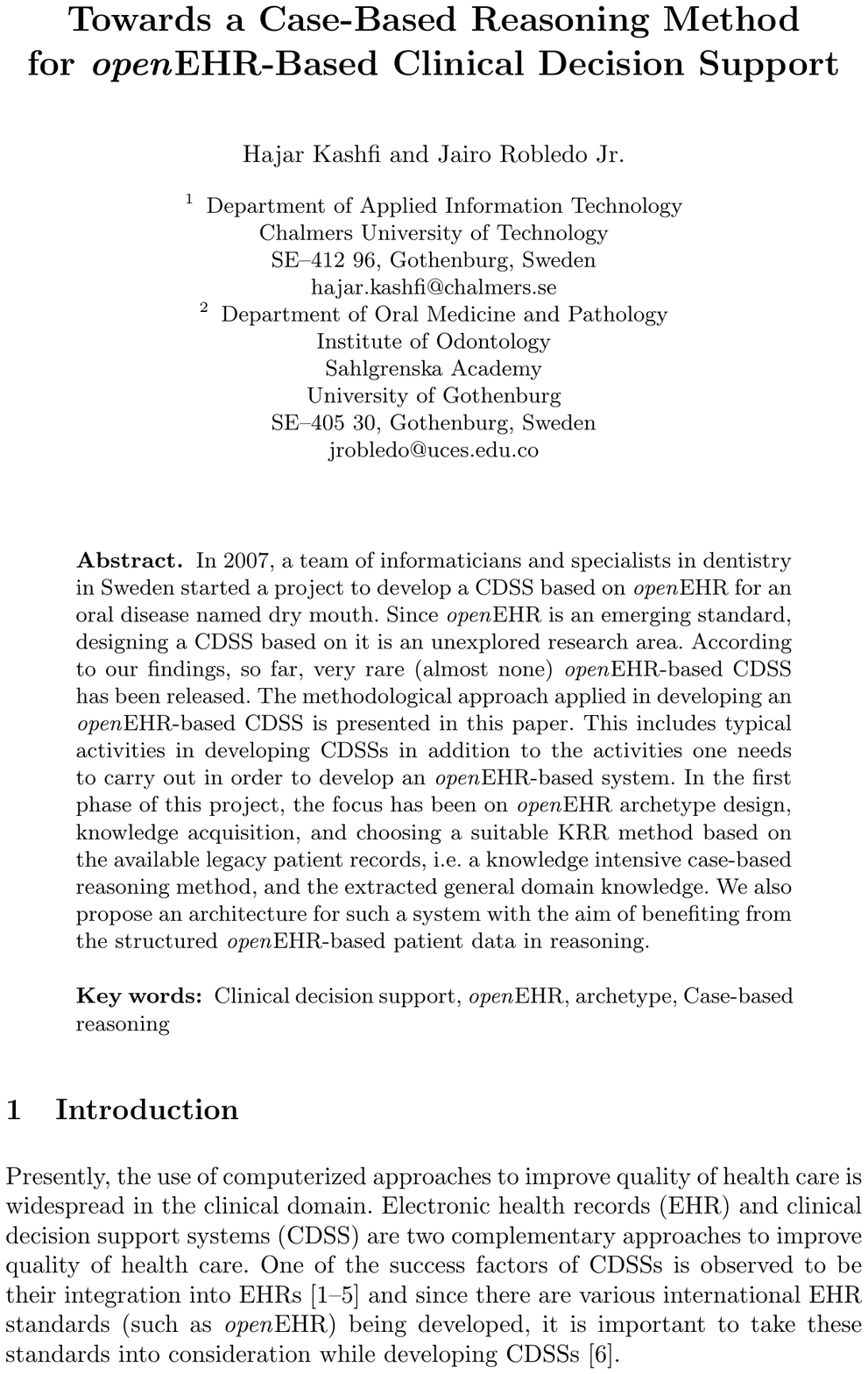}

\newpage
\vspace*{3cm}
\thispagestyle{empty}
\end{document}